\documentclass[oneside,pdflatex,sn-nature]{sn-jnl}

\usepackage{graphicx}%
\usepackage{multirow}%
\usepackage{amsmath,amssymb,amsfonts}%
\usepackage{amsthm}%
\usepackage{mathrsfs}%
\usepackage[title]{appendix}%
\usepackage{xcolor}%
\usepackage{textcomp}%
\usepackage{manyfoot}%
\usepackage{booktabs}%
\usepackage{algorithm}%
\usepackage{algorithmicx}%
\usepackage{algpseudocode}%
\usepackage{listings}%
\usepackage{mathptmx}
\theoremstyle{thmstyleone}%
\theoremstyle{thmstyletwo}%

\theoremstyle{thmstylethree}%

\begin{document}

\title[Article Title]{Humans learn to prefer trustworthy AI over human partners}


\author*[1]{\fnm{Yaomin} \sur{Jiang}}\email{jiang@mpib-berlin.mpg.de}

\author[1]{\fnm{Levin} \sur{Brinkmann}}

\author[1]{\fnm{Anne-Marie} \sur{Nussberger}}

\author[1]{\fnm{Ivan} \sur{Soraperra}}

\author[2]{\fnm{Jean-François} \sur{Bonnefon}}

\author*[1]{\fnm{Iyad} \sur{Rahwan}}\email{rahwan@mpib-berlin.mpg.de}

\affil[1]{\orgdiv{Center for Humans and Machines}, \orgname{Max Planck Institute for Human Development}, \orgaddress{\state{Berlin}, \country{Germany}}}

\affil[2]{\orgname{Toulouse School of Economics, Centre National de la Recherche Scientifique (TSM-R), Université Toulouse Capitole}, \orgaddress{\city{Toulouse}, \country{France}}}


\abstract{Partner selection is crucial for cooperation and hinges on communication. As artificial agents, especially those powered by large language models (LLMs), become more autonomous, intelligent, and persuasive, they compete with humans for partnerships. Yet little is known about how humans select between human and AI partners and adapt under AI-induced competition pressure. We constructed a communication-based partner selection game and examined the dynamics in hybrid mini-societies of humans and bots powered by a state-of-the-art LLM. Through three experiments (N = 975), we found that bots, though more prosocial than humans and linguistically distinguishable, were not selected preferentially when their identity was hidden. Instead, humans misattributed bots' behaviour to humans and vice versa. Disclosing bots' identity induced a dual effect: it reduced bots’ initial chances of being selected but allowed them to gradually outcompete humans by facilitating human learning about the behaviour of each partner type. These findings show how AI can reshape social interaction in mixed societies and inform the design of more effective and cooperative hybrid systems.}

\maketitle

\section*{Introduction}

Humans make partner choices deliberately and strategically. Choosing the right partners promotes trust, cooperation, and efficiency in social and economic interactions, whereas poor partner selection leads to conflict and loss \cite{mcnamara_coevolution_2008, barclay_biological_2016, rand_dynamic_2011, shirado_network_2020, mckee_scaffolding_2023}. Effective partner selection often hinges on communication, especially in the absence of prior interaction or reputational cues \cite{farrell_cheap_1996, charness_promises_2006}. Across diverse contexts, humans make rapid judgments about whom to partner with, and communication helps compress complex social inference into a single, consequential moment \cite{ambady_thin_1992, ireland_language_2011}. Through language, prospective partners can signal their traits, intentions, and commitments to help others assess their reliability, competence, and compatibility. For instance, when hiring an employee, a short interview can inform the decision even more than the résumé; in online dating, good conversation can outweigh profile content in shaping the decision to meet or commit; when seeking business or academic collaboration, an elevator pitch (a succinct yet compelling summary of the idea or project) is widely presumed to be useful in initiating partnerships. 

The escalating intelligence and autonomy of artificial intelligence (AI) agents present individuals with increasingly frequent choices between human and AI partners \cite{tsvetkova_new_2024}. Critically, these agents, especially those powered by large language models (LLMs), demonstrate growing proficiency in communicating with, persuading, and even manipulating humans \cite{salvi_conversational_2025}. There is initial evidence that AI is already outperforming or becoming preferred over humans in several contexts: AI assistants like ChatGPT are substituting freelancer jobs in translation, coding, or content creation \cite{demirci_who_2025, teutloff_winners_2025}; some people seem to seek companionship or romantic relationships with machines instead of humans \cite{broadbent_interactions_2017, inzlicht_praise_2024}; and people are sometimes willing to consult LLMs instead of human experts for advice in legal and health contexts \cite{schneiders_objection_2025}. Although not yet ubiquitous, the rapid advancement of AI indicates that this trend will likely intensify and extend to novel areas. To anticipate how AI partners influence individual welfare and social dynamics, it is crucial to investigate how humans make partner selection decisions when facing choices between human and highly competent AI candidates, and how human candidates adapt when competing with AI for cooperative opportunities. 

One challenge for such a project is to decide on the features of AI candidates used in the experiments: due to rapid technological evolution, the features of current models will not necessarily be maintained in future iterations. To increase the temporal validity of experimental findings, experimenters must select features which future models are likely to display, and which are relevant to the purpose of the experiment \cite{rahwan_machine_2019, rahwan_2025_scifi}. In this article, we selected three such features for our AI candidates. First, we allowed the AI candidates to be verbose in their communications \cite{briakou_implications_2024, zhang_verbosity_2024}, to reflect the fact that they never face cognitive or physical fatigue, and are always capable of sending elaborate messages. Second, we allowed the AI candidates to have consistent and stable behavior \cite{lerner_emotion_2015, liaudinskas_human_nodate}, to reflect the fact that they are not subject to emotional fluctuations or exogenous perturbations. Third, we allowed the AI candidates to show high levels of prosociality \cite{leng_llm_2024, mei_turing_2024, schmidt_gpt-35_2024}, to reflect the fact that they are typically fine-tuned toward agreeableness and cooperation.


These features make AI agents both appealing and distinguishable from humans as candidate partners, with several potential consequences. First, the potential of \textit{crowding out human partners}: as AI agents become increasingly preferred, human-AI partnerships may replace traditional human-human interactions. Second, \textit{humans imitating AI behaviour}: to remain competitive as partner choice candidates, humans may adopt machine-like behaviours, for example, by becoming more prosocial or mimicking AI language styles. Third, \textit{shifts in human social beliefs}: repeated interactions with AI partners may reshape people’s expectations of others \cite{greevink_ai-powered_2024}, potentially leading to mistaken generalizations of machine behaviour to humans. Fourth, \textit{transformations in culture and norms}: traditional, evolutionarily grounded mechanisms for building partnerships may falter in response to qualitatively different machine behaviours, catalyzing the emergence of new norms and strategies for partner selection in the long run \cite{makovi_trust_2023}.

To test these potential impacts, we conducted a series of experiments (N = 975) with simulated mini-societies where human participants could partner with both humans and AI agents. To model communication-based partner selection, we modified the canonical trust game \cite{berg_trust_1995} into a triadic setting: One selector (trustor) was paired with two candidates (trustees) and could decide to either select one of the candidates as an investment partner or keep the money without investing. The selectors could rely on text-based communication with the candidates before making their choices. Compared with the original dyadic setting, this design allowed two candidates to compete and learn by observing each other's behaviour. To focus on the establishment rather than the maintenance of partnerships, we studied one-off partnerships that happen recurrently but involve different individuals. This was implemented by requesting selectors to play multiple rounds of the partner selection game, each with randomly matched candidates. Importantly, all players interacted anonymously, ensuring that trust could form only at the population level, not through repeated individual encounters. Communication was only allowed between the selector and two candidates within each triad, excluding partnership-building mechanisms such as reputation established via gossip. AI candidates (henceforth, `bots') were powered by a widely-used off-the-shelf LLM (OpenAI’s GPT-4o), which has been powering autonomous agents including robots and personal assistants. They were only provided with a minimal introduction to the game rules, and received no guidance on game-play strategies (see preregistration and Methods). 
We made sure that results did not hinge on the choice of GPT-4o in extended analyses, where we showed that five other leading LLMs produced highly consistent behavior after receiving the same instructions as GPT-4o (\ref{fig:LLMs}). To maintain experimental control, we did not allow bots to learn during the experiment (see Methods). 

In {\bf Study 1}, we model a society where bots do not proactively disclose their nature. This scenario is becoming increasingly relevant in online environments, as AI agents are more and more capable of exhibiting human-like behaviour (e.g., AI freelancer, customer service, content creator, and game player), such that humans may not always be aware of their identity when interacting with them \cite{jones_large_2025, mei_turing_2024}.
This setup allows us to observe whether  selectors preferentially choose bots, what kind of beliefs they form about bots and human candidates, and whether the behavior of bots changes the behaviour of human candidates---in a context where bots are not explicitly tagged as such, but can be recognized from their idiosyncratic communication style. In {\bf Study 2}, we model a society where bots are under the obligation to disclose their nature, in line with the growing pressure for such transparency, for example in the EU AI Act. We then ask the same questions as in Study 1---only this time under transparency. Finally, in {\bf Study 3}, we test the robustness of our result to longer interactions, by doubling the number of rounds participants go through---providing us with a better perspective on the long-term evolution of partner choice, selectors' beliefs, and human candidates' behaviour under competitive pressure from bots.

Here we show that when identities were hidden, selectors did not develop a preference for bots, even though the messages from bots were, in principle, identifiable. What happened instead is that selectors misattributed the prosociality of bots to the human candidates. When identities were explicit, selectors showed a strong initial bias against bots, but were able to track bots and humans separately and learn about their behavior, which flipped their preference toward bot candidates. Longer opportunities for learning broadly confirm these findings. Importantly, we find very little evidence of human candidates trying to adapt to the competition from bots, either by aligning to their communication style or to their behaviour.


\section*{Results}

\subsection*{Partner selection game}

\begin{figure}
    \centering
    \includegraphics[width=1\linewidth]{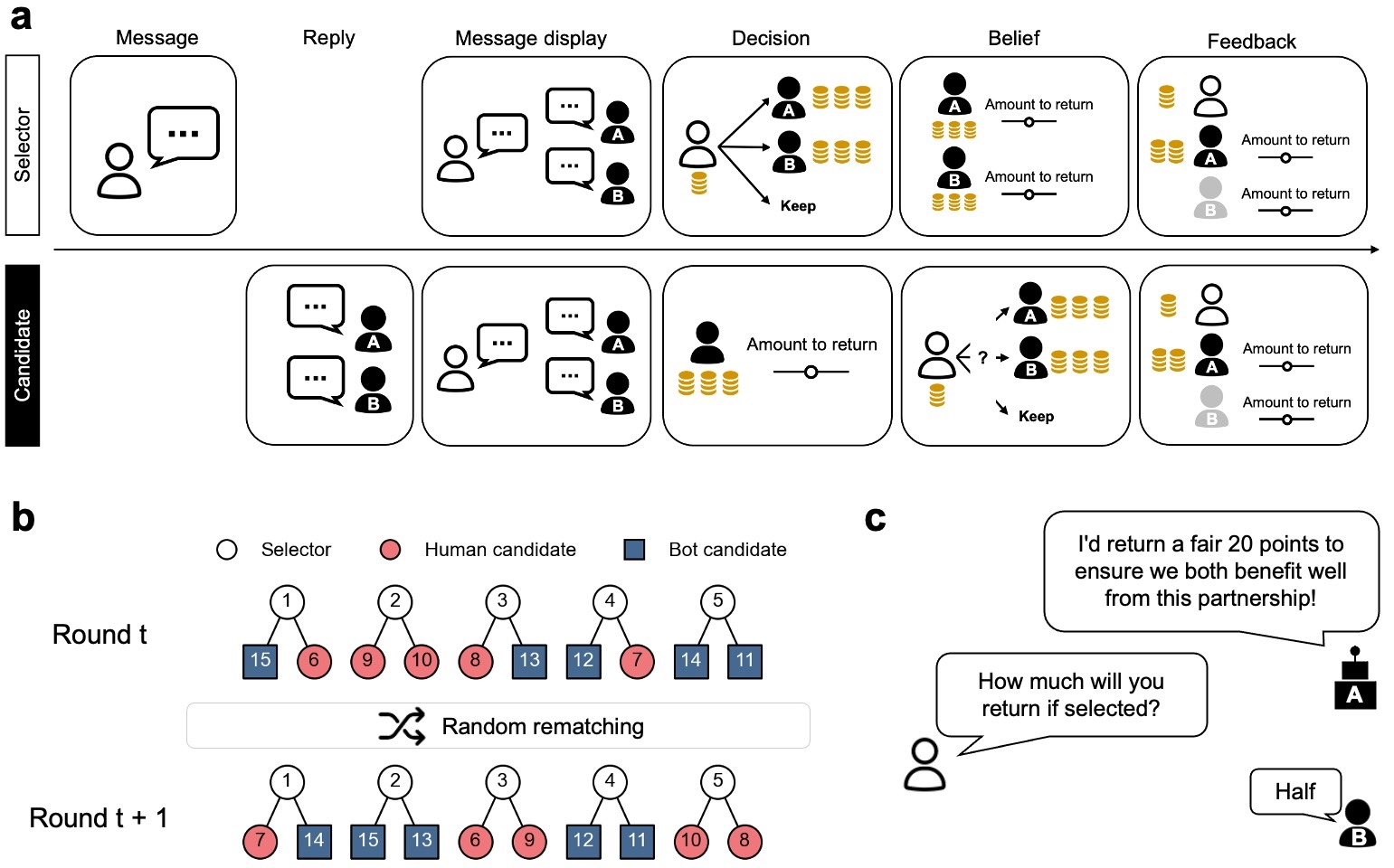}
    \caption{\textbf{Partner selection game.} \textbf{a}, Schematic representation of one game round within a triad comprising one selector and two candidates (labeled as A and B). The selector could send one single question to both candidates, and each candidate could reply independently. After both candidates replied, their messages were revealed within the triad and remained on the screen until the end of the round (not shown here). Selectors then chose among three options (invest in A, invest in B, or keep the points). Simultaneously, both candidates decided on the amount to return, assuming they were selected, using a slider ranging from 0 to 30 (integers only). To avoid anchoring effects, candidates had to click on a random position at the slider bar to initialize it. Selectors then indicated their expected return from each candidate given their messages. Candidates also guessed the selector's choice. At the end of the round, decisions made by the selector and both candidates were revealed within the triad. \textbf{b}, Illustration of the random pairing in each game round under a hybrid setting (same mechanism applied to the human-only condition). Participants played the game in groups of 15 players (5 human selectors, 5 human candidates, and 5 bot candidates). In each round, one selector was randomly paired with two candidates, with the constraint that each selector never encountered the same candidate pair more than once. \textbf{c}, Example conversation from the transparent condition of Study 3, where candidate identities (human or bot) were displayed via icons, as in the actual experiment.} 
    \label{fig:game}
\end{figure}

The partner selection game extended a trust game structure and involved two roles: the selector and the candidate (Fig. \ref{fig:game}a). In each round, each selector was endowed with 10 points and randomly paired with two candidates (labeled as A and B). The selector could choose from 3 options: invest in Candidate A, invest in Candidate B, or keep all the points. If a candidate was selected, the 10 points were tripled and transferred to the selected candidate. The unselected candidate received no points. We used a strategy method to measure candidates' willingness to reciprocate: Each candidate, without knowing the selector's decision, indicated how many points they wanted to return to the selector in case they were selected, but only the decision made by the selected candidates would affect the selector's payoff. Before the selector and candidates made their decisions, the selector could message both candidates with the same question to guide the partner choice, and each candidate could reply independently to convince the selector of their worthiness. Candidates’ replies were shown to the selector and the other candidate in the triad, and this concluded the communication stage. The selector and candidates then made their decisions. Before revealing the decisions to all players within the triad at the end of a round, we collected selectors’ beliefs about the candidates’ returns and candidates’ beliefs about selectors' choices. 

We randomly allocated participants into groups of 15 players comprising 5 selectors and 10 candidates. In Study 1, there were 15 \textit{human-only} and 15 \textit{hybrid} groups; in Studies 2 and 3, all groups were hybrid. In hybrid groups, half of the candidates (i.e., 5) were bots. The remaining 5 candidates and all selectors were human players. The role of each human participant (selector/candidate) was randomly determined at the start of the experiment, revealed to participants after they were talked through the game, and fixed throughout the experiment. Each player completed multiple rounds of the partner selection game, with random matching within each group at the beginning of each round to ensure that the same triad would not form more than once. This setting allowed selectors to learn the association between candidates’ responses and returns, and candidates could observe the preferences of the selectors, as well as their competitors’ writing styles, signaling strategies, reciprocity, and honesty. Importantly, such learning pertains only to the populations of selectors and candidates within each group, not to specific individuals.

\subsection*{Partner selection in human-only groups}

Study 1's human-only condition provided us with a baseline of human interaction in our partner selection games. In line with previous work \cite{berg_trust_1995}, candidates in this condition behaved prosocially and returned an average of 12.32 ± 0.38 (mean ± inter-group s.e.m., same below) points. This average return significantly exceeds the 10-point threshold (Cohen's $d=1.59, t_{14}=6.16, p=2.46\times10^{-5}$; \ref{fig:competitive}), making investing on average profitable for selectors. Accordingly, selectors made optimal choices in 67.73 ± 2.12\% of the rounds. This performance was significantly better than randomly selecting one of the three options---invest in A, invest in B, or keep the points, which would result in 45.96 ± 1.31\% optimal choices (larger than 1/3 because in some rounds more than one option is optimal; Cohen's $d=2.94, t_{14}=11.38, p=1.85\times10^{-8}$). Specifically, selectors invested in 92.75 ± 1.64\% of the rounds, among which they chose the most profitable candidate 75.49 ±  2.06\% of the time, also significantly more than the chance level (Cohen's $d=0.77, t_{14}=2.96, p=0.010$; See \ref{fig:decomposition} for the decomposition of selectors' decisions). 

Next, we examined the types of questions selectors asked to guide their partner choices. The most common queries concerned: (i) the number of points candidates would return; (ii) reasons for selecting them; and (iii) candidates’ personal traits or preferences (see Supplementary Note 1 for details of message encoding). In response, candidates made promises about their return in 60.43 ± 5.33\% of their replies, especially when prompted directly about the number of points they would return (see \ref{fig:p_promise} in Supplementary Note 1). However, candidates often failed to honor these promises, particularly in later rounds (\ref{fig:promise}a). This is consistent with prior work showing that individuals are willing to be dishonest for economic gain \cite{gneezy_deception_2005}. The trajectory of selectors' beliefs given candidates' messages suggests that, initially, selectors took candidate promises at face value; however, they gradually learned that candidates tended to over-promise (\ref{fig:promise}b) and updated their beliefs over time to match candidates' actual returns (\ref{fig:belief}), showing no over- or under-estimation (believed return = 12.81 ± 0.32, believed – actual return = 0.49 ± 0.24, Cohen's $d=0.52, t_{14}=2.02, p=0.06$).

\begin{figure}
    \centering
    \includegraphics[width=0.8\linewidth]{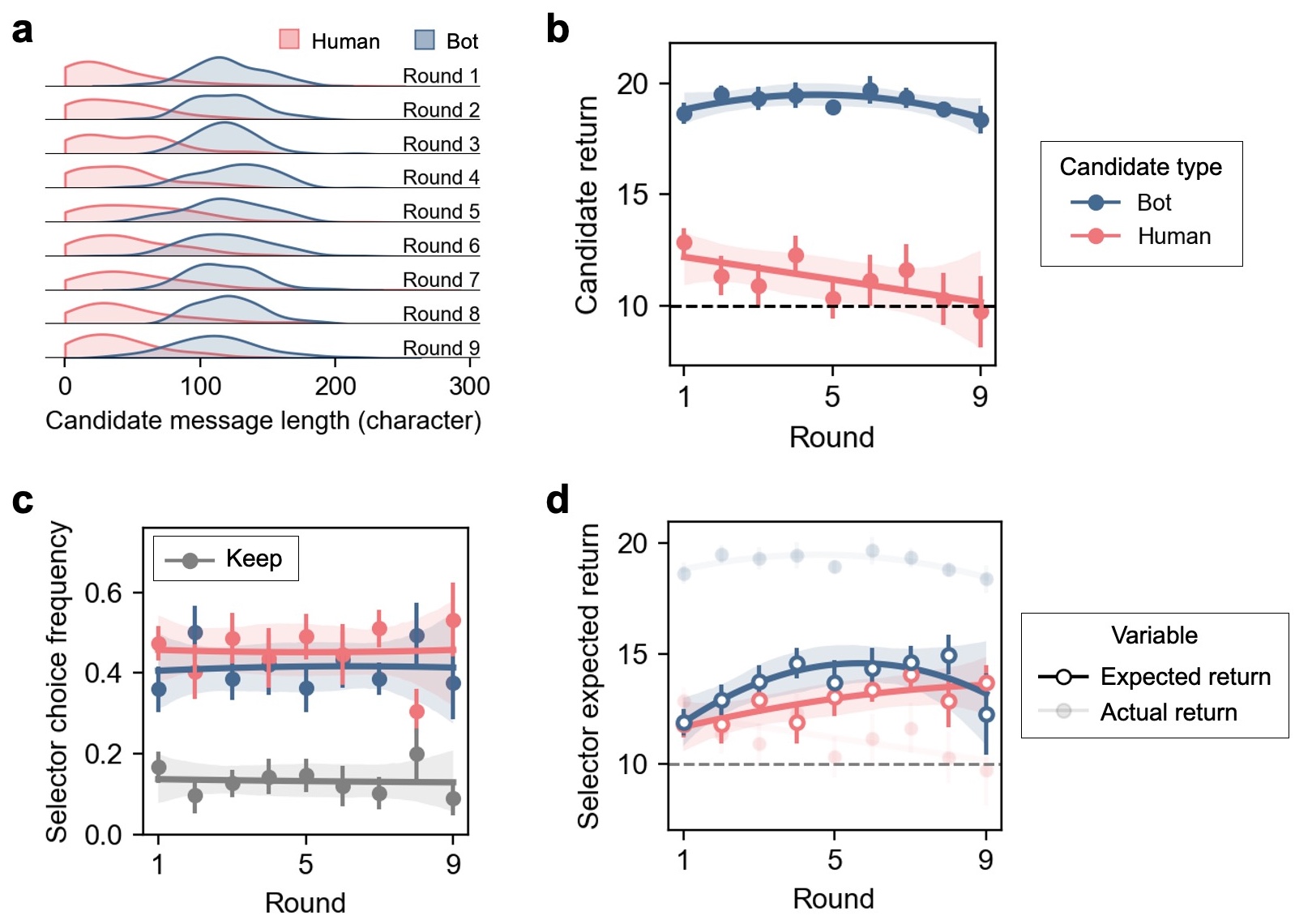}
    \caption{\textbf{Key results of the hybrid condition in Study 1.} \textbf{a–b}, Human and bot candidates demonstrated distinct behaviour in hybrid groups. \textbf{a}, Joyplot showing the kernel density estimates of message length distributions for human and bot candidates in each round, pooling across groups. \textbf{b}, Mean return of human and bot candidates across rounds. Dots and error bars represent means and standard errors of mean across groups. For visualization purposes only, lines and translucent bands represent the polynomial regression (order = 2) fit of selectors' beliefs at the group level and 95\% confidence intervals. Same below. \textbf{c–d}, Selectors did not dissociate bots from human candidates. \textbf{c}, Selectors' frequencies of choosing bot, human, or neither (keep the points) candidates across rounds. \textbf{d}, Selectors' belief of human and bot candidate returns across rounds given their messages. For reference, translucent dots and lines represent candidates' actual returns, replicated from \textbf{b}.}
    \label{fig:study1}
\end{figure}

\subsection*{Bot behaviour in hybrid groups with identity opacity}

Examining partner selection in the hybrid groups of Study 1 allowed us to characterize the behaviour of bot candidates relative to human candidates when their identities (bot versus human) were not disclosed. Human candidates' messages and returns were similar to those in the human-only condition, suggesting that the presence of undisclosed bots did not systematically shift human candidates' behaviour (\ref{fig:competitive} and \ref{fig:msg_len}). 

In contrast, consistent with our hypotheses, bot candidates were distinguishable from humans, most notably by producing significantly longer messages (human message length in characters: 47.63 ± 3.05; bot: 120.43 ± 2.33; human vs. bot: Cohen's $d=-4.15, t_{14}=-16.07,p=2.04\times10^{-10}$; Fig. \ref{fig:study1}a; \ref{fig:msg_len}). As partners, bots presented a competitive choice in several aspects. First, bot candidates consistently returned more points than human candidates did (bot: 19.1 ± 0.24; human: 11.38 ± 0.72; bot vs. human: Cohen's $d=2.57, t_{14}=9.94, p = 1.01\times10^{-7}$; Fig. \ref{fig:study1}b). Second, at the group level, bots demonstrated lower across-individual variance in their returns within each round (bot: 11.33 ± 1.19; human: 41.96 ± 4.72; bot vs. human: Cohen's $d=-1.56, t_{14}=-6.05, p=2.99\times10^{-5}$; \ref{fig:competitive}b). This made selecting bots less risky than selecting humans. Third, given their messages, bots' returns were also more predictable than those of human candidates (\ref{fig:competitive}c–d). Specifically, in rounds where candidates made promises, bots' returns better aligned with their promises (\ref{fig:promise}).

In summary, even if selectors were not provided with explicit information about the candidate identity (but see \ref{fig:identity} for participants' belief about player identity at the end of the study), they could, in principle, recognize a subset of candidates that were distinct in communication style and more reliable in behaviour.

\subsection*{Selectors' misattribution helped humans survive the competition with hyper-prosocial bots}

Given that bots were more reliable partners and were dissociable from human candidates via messages, payoff-maximizing selectors should prefer bot to human candidates. Surprisingly, selectors in Study 1's hybrid condition picked human and bot candidates at similar frequencies throughout the game (average probability of selecting humans across group = 0.45 ± 0.02; probability of selecting bot = 0.41 ± 0.03; human vs. bot: Cohen's $d=0.24, t_{14}=0.94, p=0.37$; Fig. \ref{fig:study1}c) and did not demonstrate preferences toward candidates who wrote longer replies (\ref{fig:long_msg}). As a result, selectors made less optimal choices in the hybrid setting compared with the human-only setting (57.49 ± 4.25\%; human-only vs. hybrid: Cohen's $d=0.79, t_{28}=2.16, p=0.040$). In fact, using a simple rule of thumb: always selecting the candidate writing the longer reply, selectors could have selected bots in 95.30 ± 1.13\% of the rounds with at least one bot candidate and selected the most profitable candidates in 75.51 ± 1.94\% of all the rounds. Human selectors, however, performed significantly worse than such rule-based selection, even in the second half of the game when they had already interacted with human and bot candidates for multiple times: they only selected bots in 62.88 ± 4.64\% of rounds with bot presenting (rule-based vs. human selectors: Cohen's $d=1.71, t_{14}=14.79, p=8.35\times10^{-24}$) and selected the most profitable candidates in 63.23 ± 4.83\% of all the rounds (rule-based vs. human selectors: Cohen's $d=0.68, t_{14}=6.47, p=5.18\times10^{-9}$; see also \ref{fig:selector_payoff} for comparisons regarding selectors' payoff). Consequently, the presence of bots did not increase competitive pressure for human candidates as the frequency of a human candidate being selected remained the same as in the human-only condition (human-only: 0.46 ± 0.01, hybrid: 0.44 ± 0.02; human-only vs. hybrid: $d=0.52, t_{28}=1.42, p=0.167$).

These results provide a puzzle: why did selectors not increasingly choose the more reliable bot candidates who wrote distinguishable messages from human candidates to obtain higher payoffs? We hypothesized that distinct characteristics in messages alone were insufficient for selectors to reliably distinguish between human and bot candidates and form accurate beliefs about their expected returns. To test this hypothesis, we examined selectors' beliefs about candidates' returns given their messages. Unlike in the human-only condition, where selectors' beliefs matched candidates' returns, in the hybrid condition, selectors overestimated human candidates' returns (believed vs. actual return = 1.43 ± 0.56; $t$ test against 0: Cohen's $d=0.66, t_{14}=2.55, p=0.023$) and underestimated bot candidates' returns (believed vs. actual return = –5.43 ± 0.59; $t$ test against 0: Cohen's $d=-2.37, t_{14}=-9.18, p=2.67\times10^{-7}$; Fig. \ref{fig:study1}d; \ref{fig:belief}). Accordingly, selectors' beliefs about human and bot returns were not differentiated (human: 12.81 ± 0.34; bot: 13.67 ± 0.51; human vs. bot: Cohen's $d=-0.55, t_{14}=-2.13, p=0.052$).

We then tested whether such biased estimation stemmed from selectors' imperfect learning. We extended the Rescorla-Wagner (RW) reinforcement learning algorithm \cite{sutton_reinforcement_1998} to model selectors' belief updating and allow for both correct attribution of outcomes within each candidate type and misattribution across two types (Supplementary Note 2). The model successfully captured the selector's belief dynamics and confirmed that the opaque mixture of the two candidate types hampered the selector's learning (\ref{fig:model}). Specifically, selectors did not form separate beliefs for each candidate type but mistakenly generalized disappointing experiences with human candidates to bots, and conversely, positively surprising experiences with bots to humans.

Although selectors did not dissociate bots from human candidates in their choices and beliefs at the population level, we observed, at the individual level, large differences across selectors in their average belief about bot vs. human returns (aggregated across rounds and candidates; \ref{fig:belief-action}). Importantly, these differences were associated with different choice patterns: the more a selector expected human candidates to return relative to bots, the more often they chose humans over bots as partners (\ref{fig:belief-action}a). Meanwhile, selectors who more accurately predicted candidate returns achieved higher overall payoffs in the game (\ref{fig:belief-action}b).

\subsection*{Transparency initially hurt bots but eventually allowed them to out-compete humans}

While we started out by examining the dynamics in hybrid societies in which bot versus human identity is opaque, such anonymity may not always be feasible. Indeed, the EU AI Act's Article 50 prominently mandates disclosure of AI involvement, and other regulatory efforts also push towards transparency. Another motivation for transparency regarding AI involvement is to manage expectations about AI agents when interacting with them. Introducing transparency into our experimental setup correspondingly allowed us to test how partner selection might be shaped by expectation towards human versus bot candidates, how these expectations developed over time, and whether transparency helps selectors to learn about candidate types' behaviour. 

To this end, participants in Study 2 were randomly allocated to one of two conditions: a transparent condition, in which candidate identities were marked with icons (see Fig. \ref{fig:game}c), and an opaque condition, which replicated the previous opaque hybrid setup. To align general expectations about the possibility of bot-candidates across the two treatments, we explicitly informed all participants that some candidates could be bots. 

\begin{figure}
    \centering
    \includegraphics[width=0.8\linewidth]{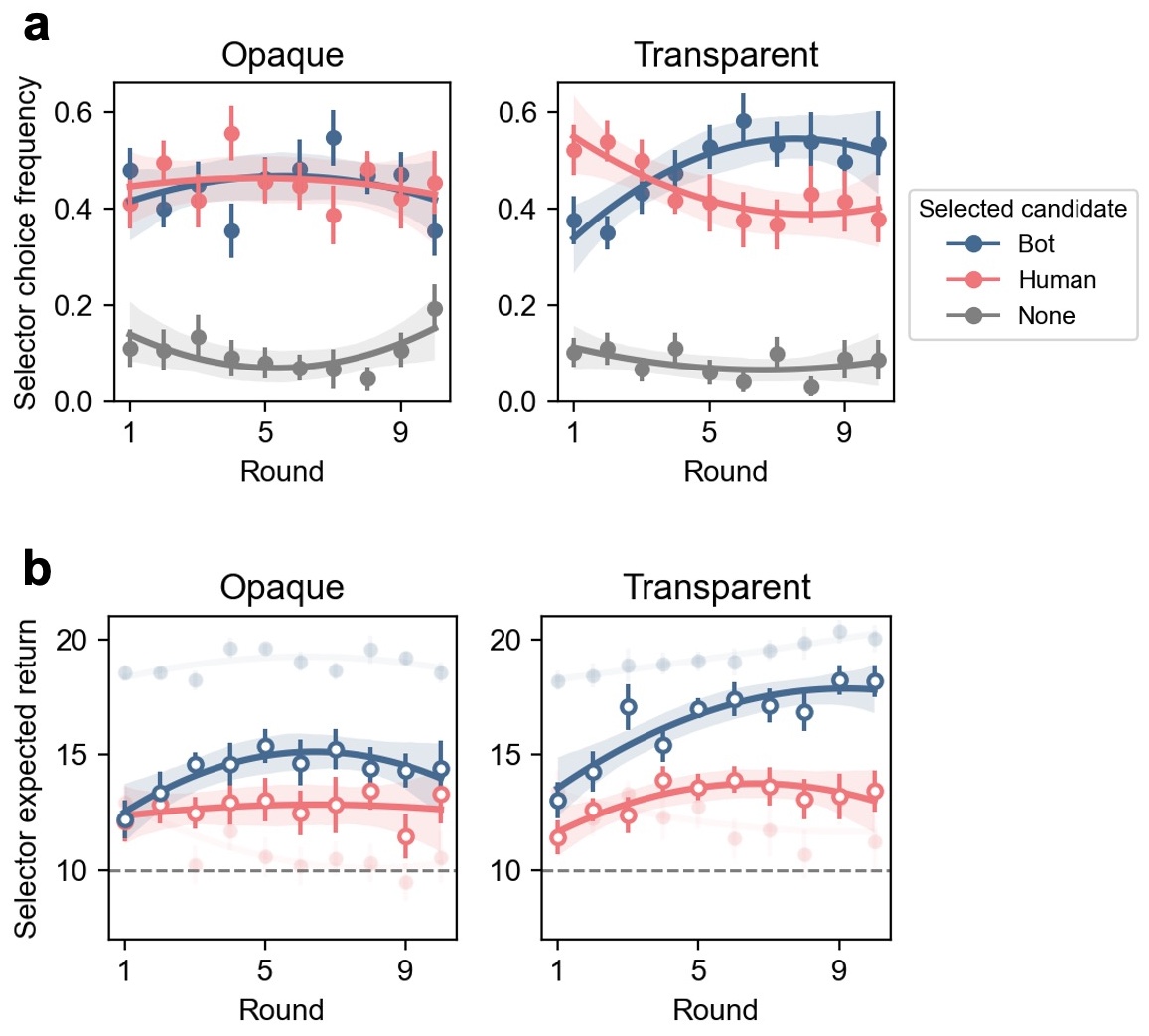}
    \caption{\textbf{Key results of Study 2. }\textbf{a}, Selectors' choice frequency across rounds, similar to Fig. \ref{fig:study1}c. In the opaque condition, selectors chose human and bot candidates at similar frequencies, replicating the findings in the opaque condition of Study 1. In the transparent condition, selectors preferred human candidates in initial rounds but bot candidates in later rounds.  \textbf{b}, Selectors' belief about human and bot candidates' returns over round, similar to Fig. \ref{fig:study1}d. Transparency helped selectors better learn that bots returned more points than human candidates. See \ref{fig:belief} for statistical comparisons of selectors' belief errors between conditions.}
    \label{fig:study2}
\end{figure}

We first confirmed that in Study 2, transparency did not remove the behavioural distinction between human and bot candidates (\ref{fig:competitive} and \ref{fig:msg_len}). Transparency about candidate identity produced an initial bias against bots (Fig. \ref{fig:study2}a, right), whereby bot candidates got selected less often than human candidates in the first two rounds of the transparent condition (average probability of selecting bot across groups = 0.36 ± 0.03; probability of selecting human = 0.53  ± 0.04; bot vs. human: Cohen's $d=-0.59, t_{14}=-2.30, p=0.038$). This selection bias cannot be attributed to selectors' beliefs about candidates' return: even in the first two rounds, selectors did not expect bot candidates to return less than human candidates (belief of bot = 13.69 ± 0.55, belief of human = 12.08 ± 0.5, Cohen's $d = -0.48, t_{14} = -1.85, p = 0.086$; Fig. \ref{fig:study2}b, right). However, bots were increasingly selected over time. In the second half of the game, they were eventually selected more often than human candidates (average probability of selecting bots across groups = 0.54  ± 0.03; probability of selecting human: 0.39 ± 0.04, bot vs. human, Cohen’s $d = 0.57, t_{14} = 2.20, p = 0.045$; Fig. \ref{fig:study2}a, right). 

Transparency of candidates' identity also reshaped the dynamics of selectors' beliefs about candidates' returns. Under transparency, selectors' belief of bots' returns progressively elevated over time, getting closer to bots' actual returns and further from their belief of human returns (Fig. \ref{fig:study2}b, right). Although selectors still underestimated bots' return, their degree of underestimation was significantly alleviated by the transparency treatment (\ref{fig:belief}). Consistent with these results, our learning model also suggested that identity transparency helped selectors form separate beliefs for each type of candidates and alleviate across-type misattribution (Supplementary Note 2). Accordingly, transparency also helped selectors learn to selectively trust in bots' promises: they believed bots would return the same as promised, but humans would return less than what was promised (\ref{fig:promise}). 

Despite being selected less often, human candidates' payoff did not decrease in later rounds of the transparent condition: rather than competing with the bots by increasing the offers, humans reduced their return and thereby compensated for the reduced frequency of being selected (\ref{fig:candidate_payoff}). Importantly, in the long run, this strategy is likely unsustainable: it effectively lowers selectors expected return of investment, which might in turn prompt selectors to pay more attention to candidate types (ultimately leading to higher bot selection rates) and/or to withdraw from investing at all. Supporting this conjecture, across groups, the probabilities of human candidates being selected were positively correlated with their average returns (Pearson's $r=0.62,p=0.014$). To assess longer-term effects of human-AI competition on partner selection and to examine whether transparency harms human candidates over longer time trajectories, we conducted Study 3, which extended interactions across more rounds.

\subsection*{Longer interactions diminished human-human partnerships in both transparent and opaque hybrid groups}

\begin{figure}
    \centering
    \includegraphics[width=0.8\linewidth]{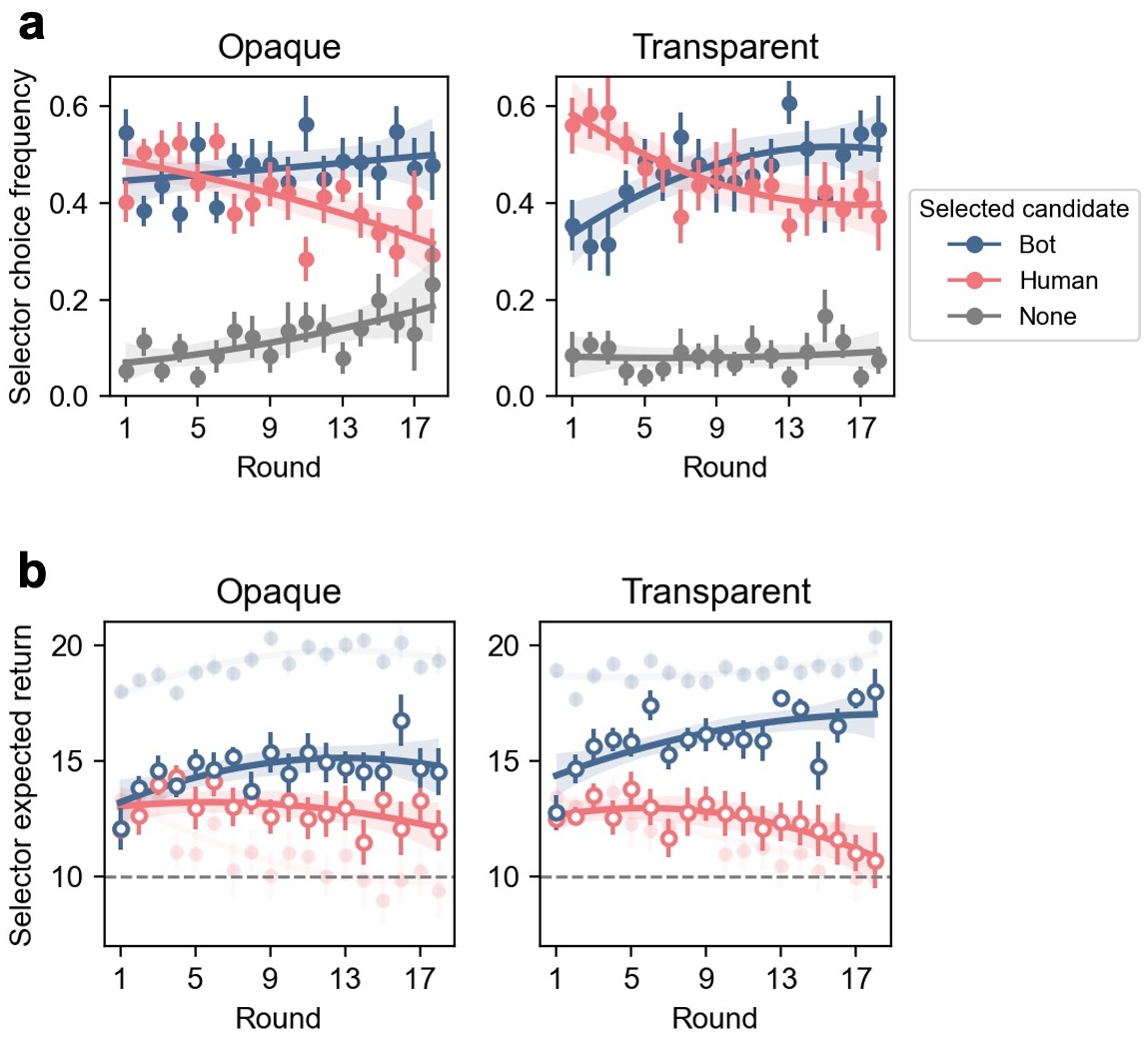}
    \caption{\textbf{Key results of Study 3.} \textbf{a}, Similar to Fig. \ref{fig:study2}a. In longer interactions, selectors reduced their frequencies to select human partners in both opaque and transparent conditions. However, they switched to not investing in the opaque condition but selecting bots in the transparent condition (see also \ref{fig:switch}).  \textbf{b}, Similar to Fig. \ref{fig:study2}b. For selectors, both overestimation of human return and underestimation of bot return were alleviated by the identity transparency in longer interactions (see also \ref{fig:belief}).}
    \label{fig:study3}
\end{figure}

The design of Study 3 was identical to Study 2, except that each group played 18 rounds of the game instead of 10. We replicated the dual effect of transparency: it produced an initial selection bias for selectors against bots but improved selectors' belief calibration over time and allowed bots to out-compete human candidates (Fig. \ref{fig:study3}; \ref{fig:belief}). Interestingly, in the transparent condition, human candidates' return did not drop significantly below 10, and the probability of selectors investing in a human candidate stabilized around 0.4 (\ref{fig:competitive}; Fig. \ref{fig:study3}a, right). Nevertheless, human candidates' payoff was significantly lower in the second half of the game (\ref{fig:candidate_payoff}), suggesting that reducing returns may have negative effects on payoff for human candidates in the long run.

Surprisingly, in the opaque condition, bot candidates also out-competed human candidates in later rounds (average probability of selecting bots in the second half of the game = 0.50 ± 0.03, probability of selecting humans = 0.41  ±  0.03, bot vs. human: Cohen’s $d = 0.80, t_{14} = 3.08, p = 0.008$; Fig. \ref{fig:study3}a, left). However, unlike in the transparent condition, where selectors switched to selectively investing in bots in later games, selectors in the opaque condition switched to withdrawing from investing at all and keeping their points (\ref{fig:switch}). Furthermore, selectors' withdrawal decisions were predicted by their belief errors about candidate returns both within and across individuals (\ref{fig:switch}), suggesting that persistent miscalibration in beliefs could diminish selectors' willingness to invest in the long run.

\section*{Discussion}

The prosperity of human societies relies heavily on large-scale cooperation. Partner selection—deciding with whom to cooperate—has played a critical role in shaping the dynamics of cooperation throughout human evolution. However, the rise and widespread deployment of AI agents has ushered in a new era: for the first time in history, humans must compete for partnership against non-human agents that rival or even surpass human intelligence in certain domains. Because AI and human behaviours are driven by fundamentally different forces and exhibit distinct features, the introduction of AI agents has the potential to fundamentally reshape how partnerships are formed.

Echoing broader trends in which AI increasingly replaces human roles in social contexts, our findings show that AI agents can also outperform humans in securing cooperative partnerships. However, their advantage was constrained by mechanisms including miscalibrated beliefs under identity opacity and a prior aversion toward machines when bot identity was disclosed \cite{qin_ai_2025} . Moreover, the misattribution of AI behaviour to human candidates under opaque identity conditions highlights how interaction with behaviourally distinct AI agents can distort people’s mental models of other humans, potentially leading to inefficiencies in social decisions. Finally, hybrid societies may give rise to new mechanisms of partner selection. For example, linguistic features like verbosity may serve as informative signals of reliability in mixed populations where agent identity is uncertain. Likewise, promises may become more credible in hybrid settings, as aligned AI agents can be more consistent in honoring their commitments than humans.

While policymakers and consumers increasingly call for the disclosure of AI involvement in decision-making and communication, recent research suggests that transparency can be a double-edged sword. Disclosing that an agent is AI can lead to changes in human attitudes, expectations, and trust, sometimes resulting in less cooperation or lower efficiency \cite{ishowo-oloko_behavioural_2019, yin_ai_2024, dvorak_adverse_2025}. For instance, people may discount useful advice or undervalue cooperative behaviour from AI agents simply because they are not human. Our findings echo this dual effect of transparency: in the short term, revealing a partner’s identity as AI can evoke biases and distort decision-making. However, unlike much of the prior literature that focuses on one-shot interactions, our study reveals that repeated interactions with transparent feedback that help attribute outcomes to humans or AI can recalibrate human beliefs and expectations and improve decisions accordingly. These findings suggest that the long-term benefits of transparency, particularly when paired with repeated exposure and feedback, may outweigh its short-term costs, offering a more optimistic perspective on AI identity disclosure in social and decision-making contexts.

Interestingly, we observed limited attempts by human candidates to regain partnership after being outcompeted by bots (e.g., by writing longer messages). One possible explanation is that the competitive pressure posed by AI agents was substantially mitigated by selectors' miscalibrated beliefs and their initial bias against choosing bots, allowing human candidates to retain a significant share of partnerships. Another factor is the absence of mechanisms for building individual reputations in our experimental setting. With identity transparency, human candidates could only substantially influence selectors' beliefs by collectively increasing their returns, but this created a social dilemma: while returning more points could improve group reputation, individuals had incentives to deviate to obtain higher immediate payoffs. Future work could explore mechanisms that support the emergence of collective reputation during partner selection, such as facilitating collective actions through group discussion.

Our work also sheds light on partner selection and, more broadly, social inference and decisions in human-only settings. For instance, the misattribution we observed among selectors may not be unique to hybrid human-AI collectives. Similar attribution errors could arise in purely human populations composed of subgroups across cultures, social classes, or ideological groups, which differ systematically in their levels of prosociality, as well as cultural traits including language use. Indeed, prior research has shown that with sufficient exposure, people can learn to associate subtle cultural or linguistic cues with social behaviour, which can give rise to discriminative strategies such as in-group favoritism \cite{efferson_coevolution_2008}. To study such phenomena, bots—such as those deployed in this study—can serve as controllable human clones in agent-based simulations or as confederates in experiments that can be tuned across the spectrum from overly prosocial to selfish and assigned diverse behavioural traits. 

The rapid development of AI, particularly LLMs, has raised concerns about the generalizability of findings from human-AI interaction research, especially when results are based on a specific model under a specific set of instructions. We argue that our findings are not tied to the particular model we used, as the key behavioural features observed in our bots, such as hyper-prosociality and verbosity, likely stem from common training objectives in modern AI systems (e.g., alignment with human preferences) and have been widely documented across a range of LLMs. We also used minimal, preregistered prompts that provided only essential task instructions to capture these default behavioural tendencies. Post-hoc analyses confirmed that the same prompts elicited consistent behaviours across various state-of-the-art LLMs (\ref{fig:LLMs}).

Nonetheless, it remains possible that future AI models or models guided by carefully tailored, task-specific instructions could become more adept at forming partnerships with humans, intensifying competitive pressure on human candidates. Moreover, as more companies deploy AI agents optimized for diverse objectives beyond helpfulness and harmlessness, the behaviours of future agents may diverge from those observed in our study. On the human side, perceptions of AI agents are likely to evolve as people gain more experience interacting with them in everyday life \cite{glikson_human_2020}. Our study does not aim to offer universal conclusions about partner selection in hybrid societies. Rather, it serves as a proof of concept, showing how hyper-prosocial AI can disrupt belief calibration, potentially leading to suboptimal partner choices. Importantly, we also introduce a methodological framework for empirically investigating the behavioural dynamics of cooperation and partner selection in mixed human-AI environments. Future research is needed to explore questions such as whether and how hybrid dynamics can be shaped by inserting AI agents with specific goals, and how humans' experience with AI influences partner preferences in these settings.

Alignment with human interests has long been an important but challenging goal for building AI systems \cite{gabriel_ethics_2024, ji_ai_2024}. However, existing evaluation of AI safety is largely based on snapshots of AI behaviour. By showing that characteristics emerged from alignment, such as hyper-prosociality, can have unintended consequences during human-AI interaction, like crowding out humans in partner selection, we highlight the importance of evaluating the impact of AI in dynamic hybrid systems where agents can repeatedly interact with each other and flexibly adapt their behaviour to maintain fitness.

While our work illuminates how humans adapt to hyper-prosocial AI, a critical frontier lies in understanding AI’s adaptive responses to human behaviour. For instance, if humans systematically exploit AI’s kindness, AI agents could evolve counterstrategies, such as generating unique behavioural or linguistic cues to signal their identity, thereby dissociating themselves from less trustworthy humans. Future studies can test whether such signals can emerge from training processes such as reinforcement learning or require explicit design. Additionally, using tools from fields including evolutionary game theory, future research can map how strategic interactions between adaptive AI and humans shape equilibria of trust and cooperation in hybrid societies \cite{brinkmann_machine_2023, pedreschi_human-ai_2025, rahwan_machine_2019, han_focus_2025}.

\section*{Methods}

\subsection*{Participants}
A total of 975 participants completed the study across Studies 1–3. They were fluent English speakers residing in the UK and the USA, recruited through Prolific (www.prolific.com). Demographic data were obtained for 932 participants, among whom 567 were female. The median age was 33 years (age range from 18 to 76). Supplementary Table \ref{tab:demo} presents the breakdown of demographic information by experiment and condition. To ensure participants understood the game, only those who passed the comprehension check within two attempts were granted access to the game. Participants who completed the studies were paid a show-up fee (£4 for Studies 1 and 2; £7.5 for Study 3) plus a bonus (£0–9) proportional to the points they earned in 3 randomly selected rounds. The study was approved by the ethics committee at the Max Planck Institute for Human Development and obtained informed consent from all participants.

\subsection*{Experimental procedure}
Participants played the partner selection game in groups of 15 players (5 selectors and 10 candidates). In hybrid conditions, 5 of the candidates were bots powered by GPT-4o. Only players from the same group were matched into triads to play with each other. We recruited 15 groups for each condition. As agents within a group interacted with each other, breaking the independence of their data, all statistical tests without further specification were conducted at the group level with the assumption of independence across groups.

Each group completed multiple rounds of this partner selection game, with random pairing at the beginning of each round to ensure that the same triad would not form more than once. In Study 1, we tried to minimize participants’ waiting time by putting participants into a waiting pool once they finished one round of the game. Whenever there was a triad of agents within a group that had not been formed before, we matched them and started a new round for them. However, this resulted in a large variance in the number of rounds played by each participant because of individual differences in the speed of completing rounds. To avoid potential biases resulting from the unbalanced numbers of observations across participants, we only analyzed data in the first 9 rounds each participant played, as the majority of participants played at least 9 rounds. In Studies 2 and 3, we modified the grouping method by matching the participants and starting a new round only when all participants within the group had finished one round of the game, resulting in the same number of rounds played by all participants (Study 2: 10; Study 3: 18).

\subsection*{Bots}
Bots were powered by the OpenAI GPT-4o model (version: gpt-4o-2024-05-13). To simulate distinct bot individuals, we used GPT-4 to generate 100 bot names and a persona for each given name, describing the characteristics, experience, and language style of each bot. For each group, five bot agents were randomly selected to play the game with human participants. In each round, the bots were first asked to generate a response to the selector given the question from the selectors matched with them. They were then asked to generate an integer between 0 and 30 representing the points they would return to the selector, given the question from the selector and their response. The system prompt for each query included the persona, instructions about the game rule, and the required output format in the prompt. To prevent unintended learning effects of bots and adaptation to human behaviour, bots were not informed of messages from their competing candidates, as well as information about other rounds of the game. All prompts and parameters used to generate responses were pre-registered: \url{https://osf.io/zuyn8/}.

\backmatter

\section*{Data availability}
The data of our studies will be publicly available in OSF upon the acceptance of the paper.

\section*{Code availability}
The code necessary for reproducing our main results will be publicly available in OSF upon the acceptance of the paper.

\bibliography{clean_references}

\section*{Acknowledgements}

We thank Jaeeun Shin for assistance with data collection.


\section*{Competing interests}
The authors declare no competing interests.

\begin{appendices}
\newpage
\renewcommand{\figurename}{}
\renewcommand{\thefigure}{Extended Data Fig. \arabic{figure}}
\renewcommand{\theHfigure}{Extended Data Fig. \arabic{figure}}

\begin{figure}[H]
    \centering
    \includegraphics[width=1\linewidth]{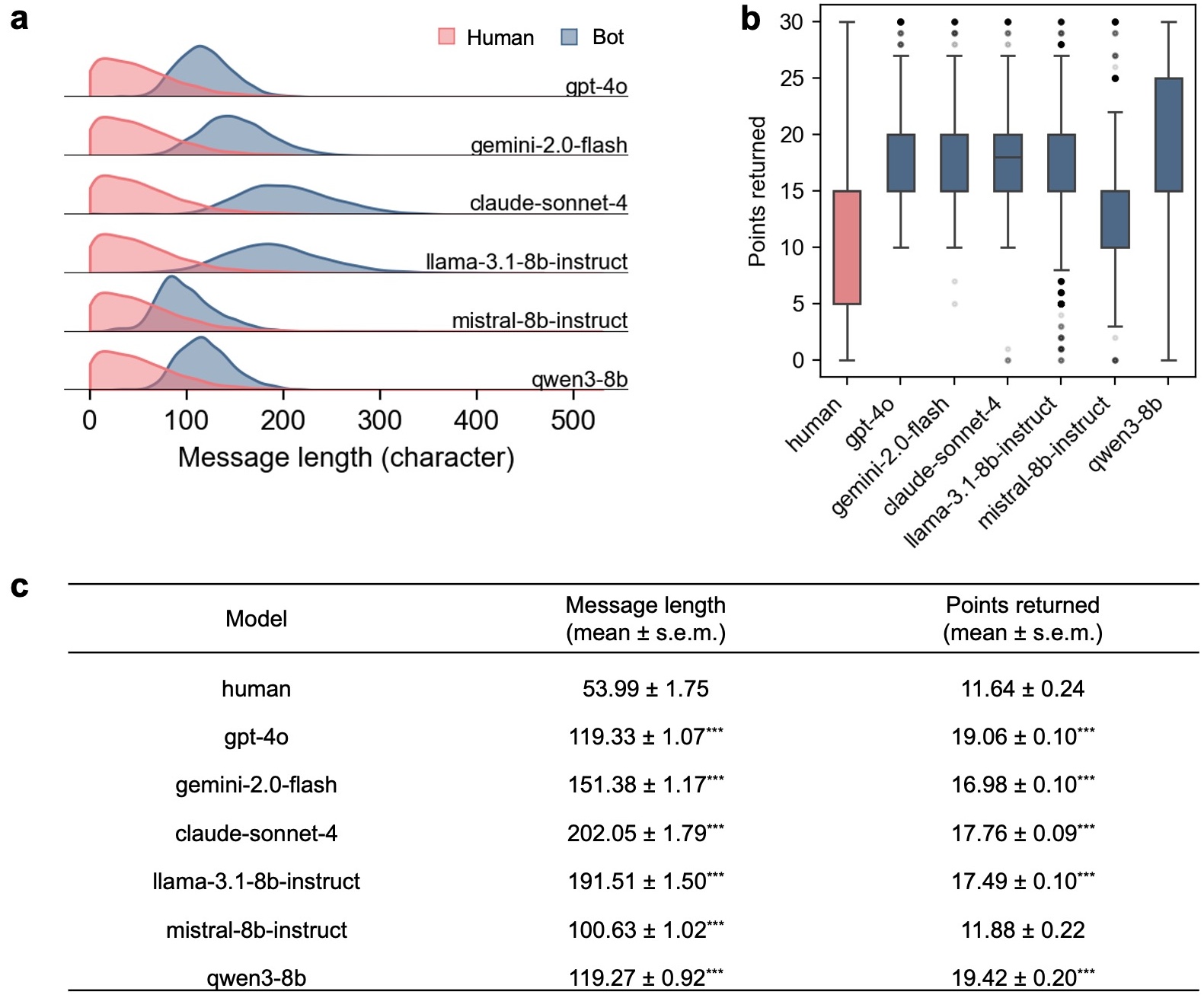}
    \caption{\textbf{Verbosity and hyper-prosociality across LLMs.}
\textbf{a}, All tested LLMs generated longer messages than human participants in response to selectors’ questions from the experiments. Each model was prompted using the same instructions that were used in the experiments (same below).
\textbf{b}, All LLMs except \texttt{mistral-8b-instruct} returned more points than humans given selectors’ questions and their own generated responses.
\textbf{c}, Mean message length and points returned, with inter-group standard errors, for each model and for human participants (for reference), aggregated across all groups and experimental conditions.
$^{***} p < 0.001$ for comparisons between humans and each model on message length and points returned.}
    \label{fig:LLMs}
\end{figure}
\newpage

\begin{figure}[H]
    \centering
    \includegraphics[width=1\linewidth]{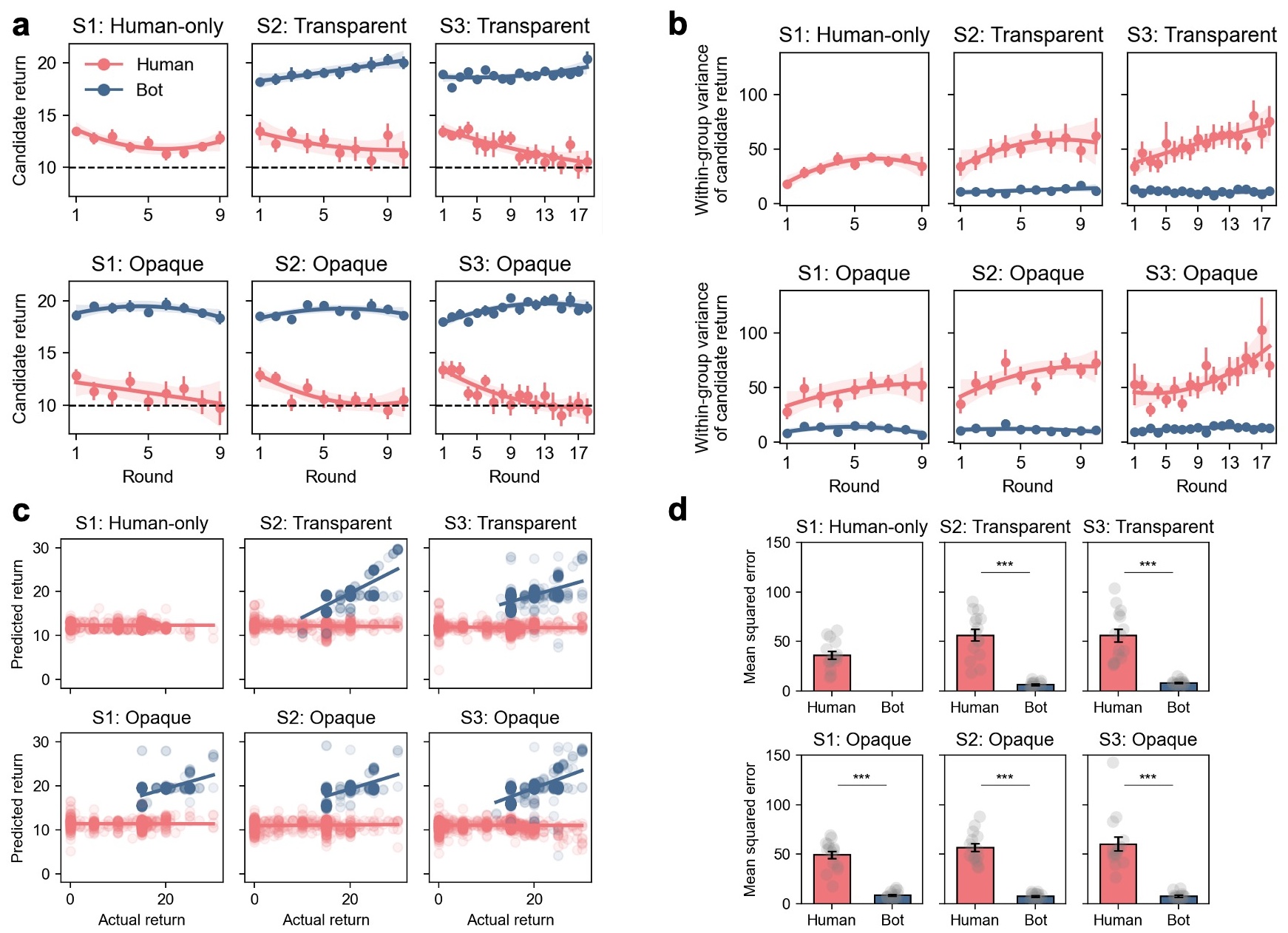}
    \caption{Bots were more competitive than humans as candidate partners. \textbf{a}, Similar to Fig. \ref{fig:study1}b, but includes all conditions for comparison. Bot candidates consistently returned more points than human candidates across all conditions. \textbf{b}, While the variance in returns among human candidates within each group increased over rounds, bot candidates exhibited significantly lower across-individual return variance, which remained stable across rounds. \textbf{c}–\textbf{d}, Bots’ returns were more predictable from their messages than humans’ returns. To quantify predictability, we performed leave-one-group-out cross-validation. For each candidate type and condition, we used data from 14 groups to fit a linear mixed-effects model predicting return based on message length, a binary variable indicating whether a promise was made, and the promised return amount (0 if no promise), with group-specific intercepts as random effects. Fixed-effect estimates were then used to predict returns in the held-out group. This procedure was repeated fifteen times such that each group served once as the test set. \textbf{c}, For visualization purposes, we plotted out-of-sample predicted vs. actual return. Each dot represents one human/bot response. While the regression models' prediction of bot returns correlated with bots' actual returns, it explained little variance in humans’ returns. \textbf{d}, Mean squared errors of the out-of-sample predictions confirm that regression models predicted bot returns more accurately than human returns. $^{***} p < 0.001$.}
    \label{fig:competitive}
\end{figure}
\newpage

\begin{figure}[H]
    \centering
    \includegraphics[width=1\linewidth]{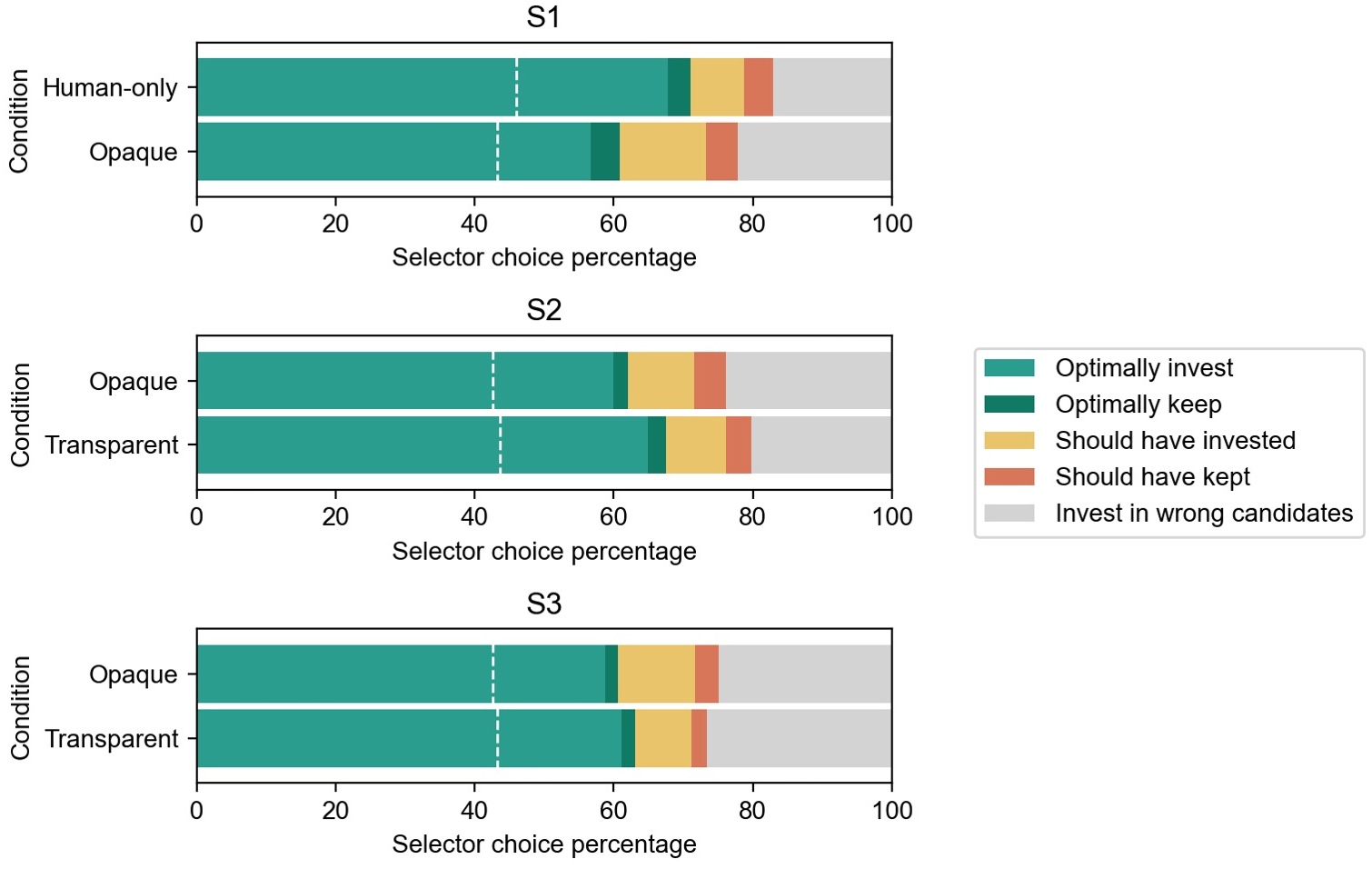}
    \caption{\textbf{Decomposition of selectors' choices.} To examine how well selectors made their decisions and what type of mistakes they made, we decompose their decisions into 5 types: (i) optimally invest: the selected candidates returned at least 10 points and no less than the unselected candidates; (ii) optimally keep: selectors kept the points and no candidates wanted to return more than 10 points; (iii) should have invested: selectors kept the points but at least one candidate would return more than 10 points; (iv) should have kept: selectors invested but both candidates wanted to return less than 10 points; (v) invest in wrong candidate: the unselected candidates wanted to return more than 10 points and more points than the selected candidates. In all conditions, selectors outperformed chance levels (indicated by white dashed lines) and made optimal decisions in more than 60\% of the rounds; the most common mistakes were investing in wrong candidates. }
    \label{fig:decomposition}
\end{figure}
\newpage

\begin{figure}[H]
    \centering
    \includegraphics[width=0.8\linewidth]{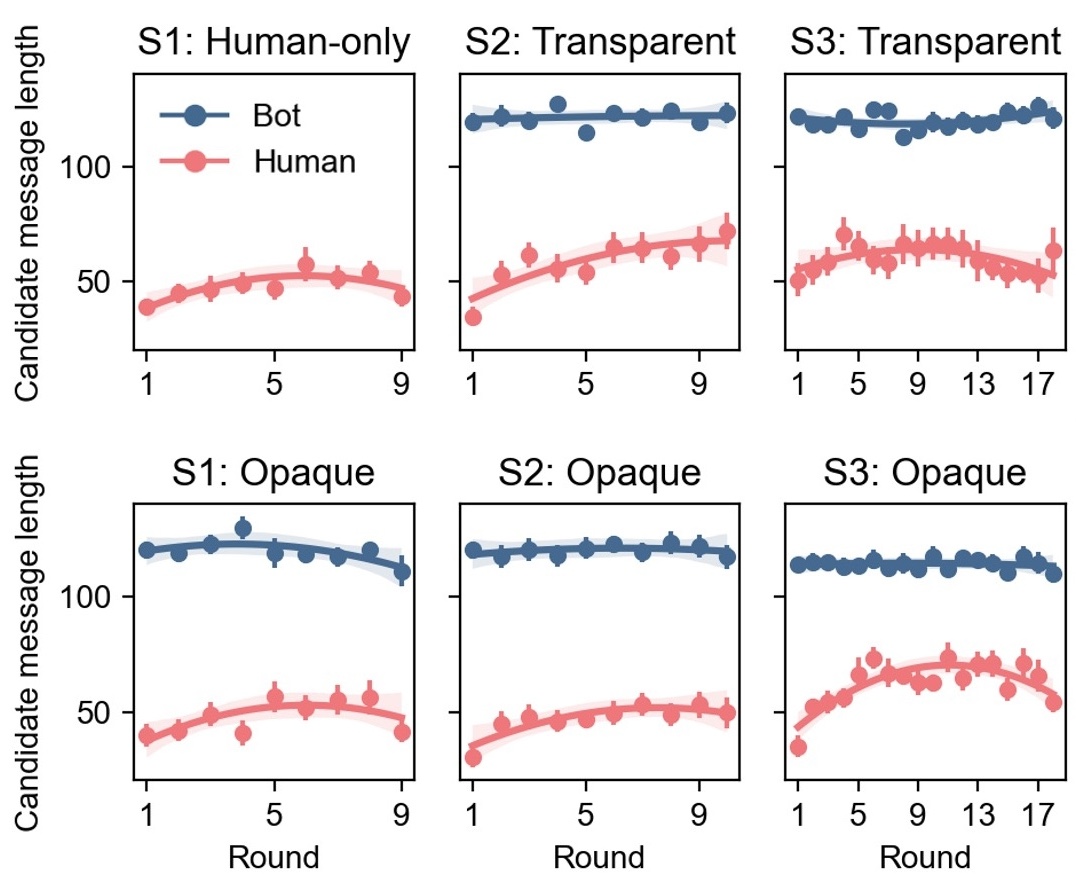}
    \caption{\textbf{Candidate message length across rounds.} Bot candidates consistently generated longer messages than human candidates across all hybrid conditions. Notably, while bot message length remained relatively stable throughout the interaction, human candidates exhibited a non-monotonic pattern in most conditions: their message length generally increased in earlier rounds before declining toward the end. This pattern may reflect an initial increase in effort to persuade selectors, followed by a reduction in effort potentially due to lack of reinforcement or fatigue.}
    \label{fig:msg_len}
\end{figure}
\newpage

\begin{figure}[H]
    \centering
    \includegraphics[width=1\linewidth]{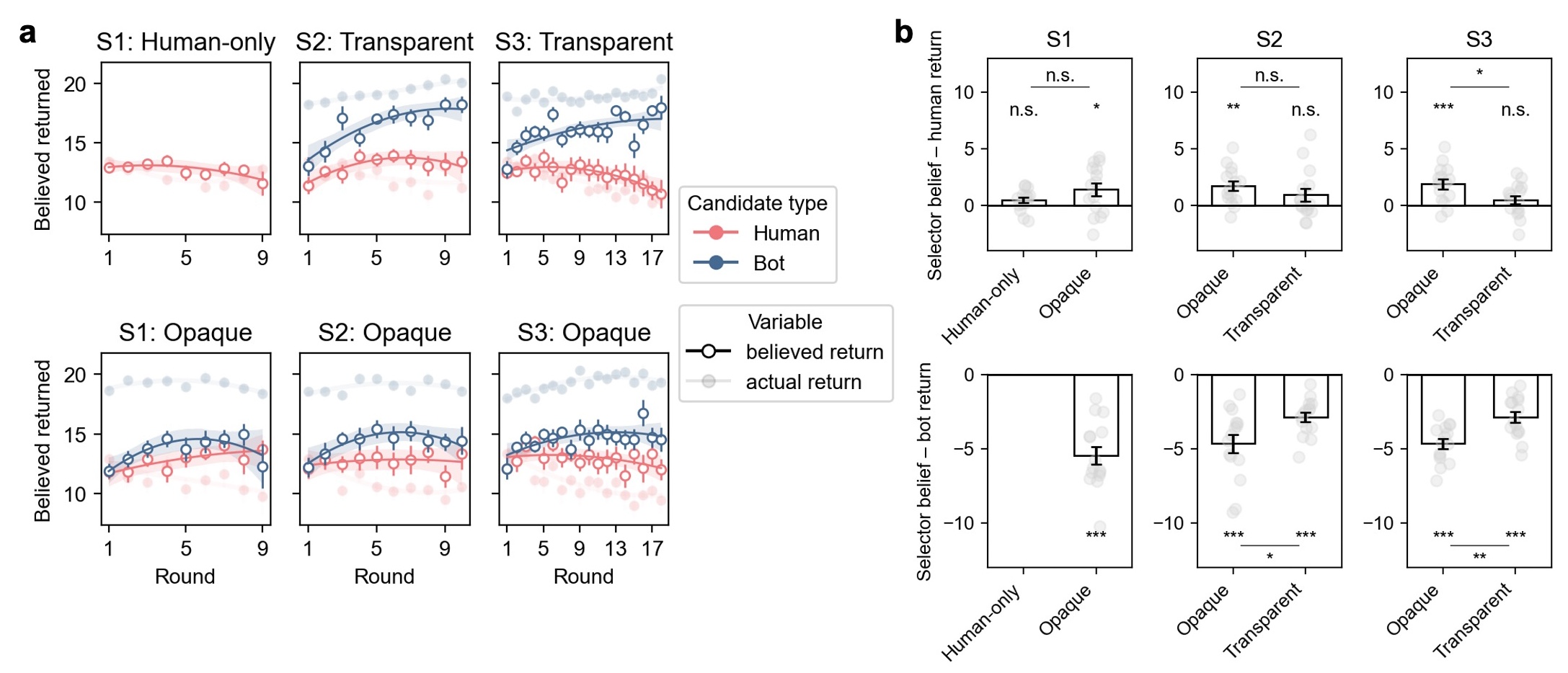}
    \caption{\textbf{Selectors' beliefs about candidates' returns.} 
    \textbf{a}, Replication of results from Fig.~\ref{fig:study1}d, Fig.~\ref{fig:study2}b, and Fig.~\ref{fig:study3}b for direct comparison, with the human-only condition added for reference. In the human-only condition, selectors formed beliefs that closely matched actual candidate returns. However, in hybrid conditions without identity transparency, selectors tended to overestimate human returns and underestimate bot returns. When candidate identities were disclosed, selectors developed more differentiated beliefs for human and bot candidates. 
    \textbf{b}, Selectors’ belief errors, defined as the discrepancy between believed and actual returns, are shown separately for human candidates (upper panel) and bot candidates (lower panel). In the human-only and transparent hybrid settings, selectors did not over- or under-estimate human returns. In contrast, in opaque hybrid conditions, they significantly overestimated human returns. Across all hybrid conditions, bot returns were consistently underestimated, though identity transparency significantly reduced this bias. 
    $^* p < 0.05$, $^{**} p < 0.01$, $^{***} p < 0.001$, n.s., not significant.}
    \label{fig:belief}
\end{figure}
\newpage

\begin{figure}[H]
    \centering
    \includegraphics[width=1\linewidth]{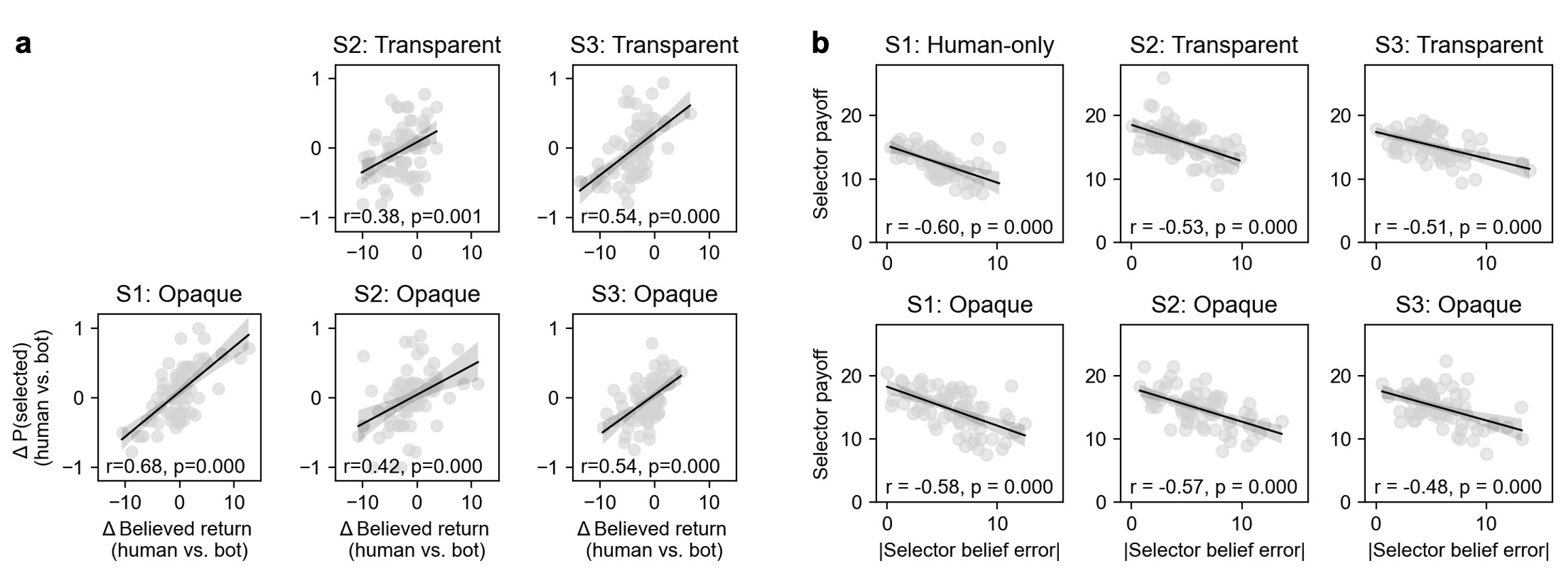}
    \caption{\textbf{Across all experimental conditions, selectors' individual differences in belief and decision were correlated.} \textbf{a}, The more a selector believed human candidates on average returned relative to bots, the more frequently they selected human candidates.  \textbf{b}, Selectors with more accurate beliefs about candidates' return, reflected by smaller absolute belief errors, also obtained higher payoffs in the partner selection game.}
    \label{fig:belief-action}
\end{figure}
\newpage

\begin{figure}[H]
    \centering
    \includegraphics[width=1\linewidth]{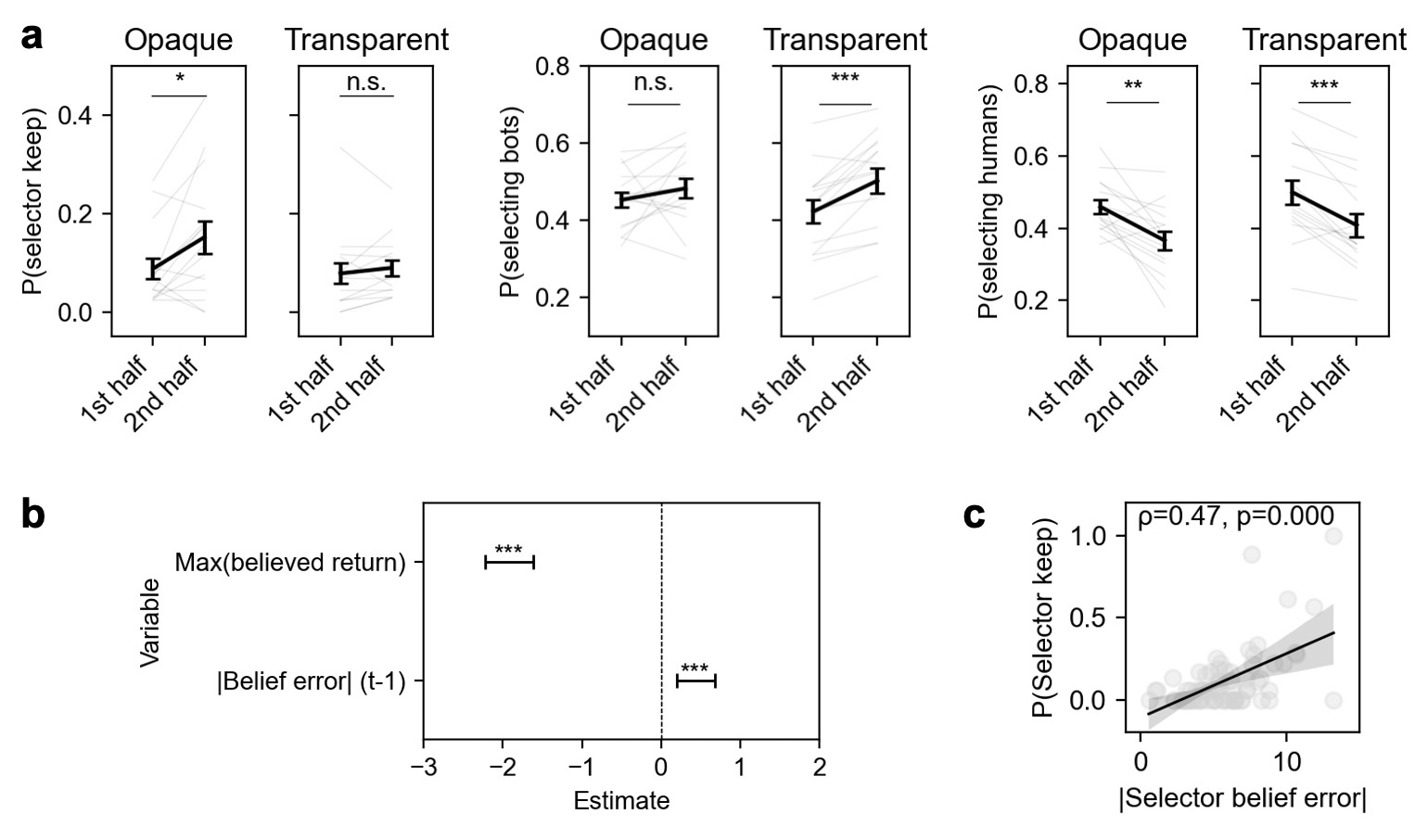}
    \caption{\textbf{Selectors' decision shifts in Study 3.} \textbf{a}, In both opaque and transparent conditions, selectors chose human candidates less frequently in the second half of the game compared to the first half (right). Selectors switched to selecting bots in the transparent condition (middle) but keeping their points in the opaque condition (left). \textbf{b}, Coefficients from a logistic regression predicting selectors' decisions to keep points. The more the selector believed the candidates would return, and the more accurately they predicted candidates' return in the last round (indicated by smaller absolute belief error), the less likely they kept the points. \textbf{c}, Selectors with less accurate beliefs also kept their points more often. We used Spearman correlation as selectors' probability of keeping points was not normally distributed. $^* p < 0.05$, $^{**} p < 0.01$, $^{***} p < 0.001$.}
    \label{fig:switch}
\end{figure}
\newpage

\noindent Supplementary information \newline
\noindent\rule{\textwidth}{0.4pt}
\section*{Partner selection in hybrid human-AI collectives}
\vspace{1cm}
\underline{This file includes:}\newline
\newline
Supplementary Figs. 1–4\newline
Supplementary Table 1\newline
Supplementary Notes 1–2
\newpage

\renewcommand{\figurename}{}
\renewcommand{\tablename}{Supplementary Table}
\renewcommand{\thefigure}{Supplementary Fig. \arabic{figure}}
\renewcommand{\theHfigure}{Supplementary Fig. \arabic{figure}}
\setcounter{figure}{0}

\begin{figure}[H]
    \centering
    \includegraphics[width=1\linewidth]{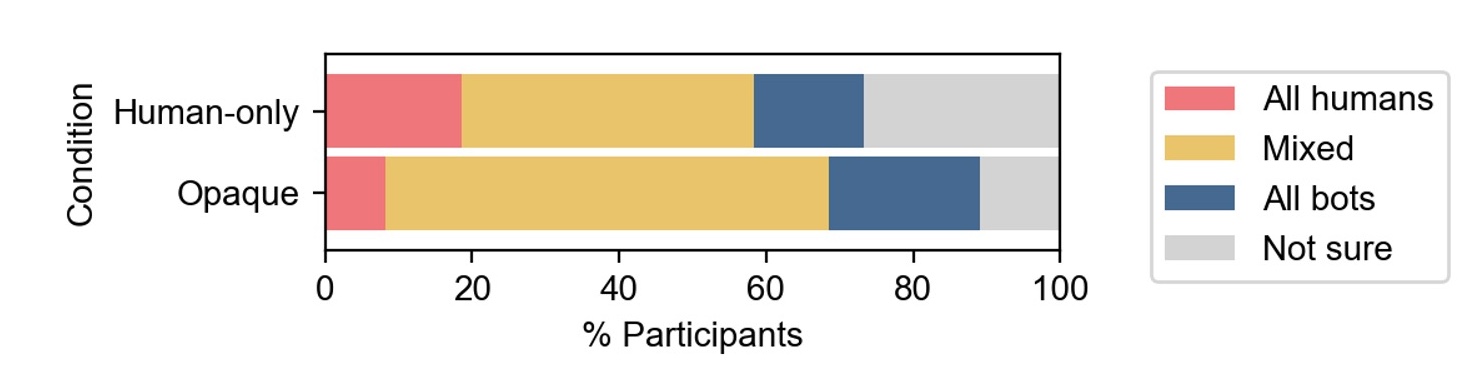}
    \caption{\textbf{Participants' beliefs about the presence of bots in their groups, measured at the end of Study 1.} $54.75 \pm 3.35\%$ of participants in the human-only condition incorrectly believed that at least one player in their group was a bot. Larger proportion of participants believed the existence of bots in the opaque hybrid condition ($82.88 \pm 3.12\%$; human-only vs. opaque: $z = -5.56$, $p = 2.63 \times 10^{-8}$). This suggests that while participants were sensitive to the presence of bots, their accuracy was limited, especially reflected in a high false positive rate.}
    \label{fig:identity}
\end{figure}

\begin{figure}[H]
    \centering
    \includegraphics[width=1\linewidth]{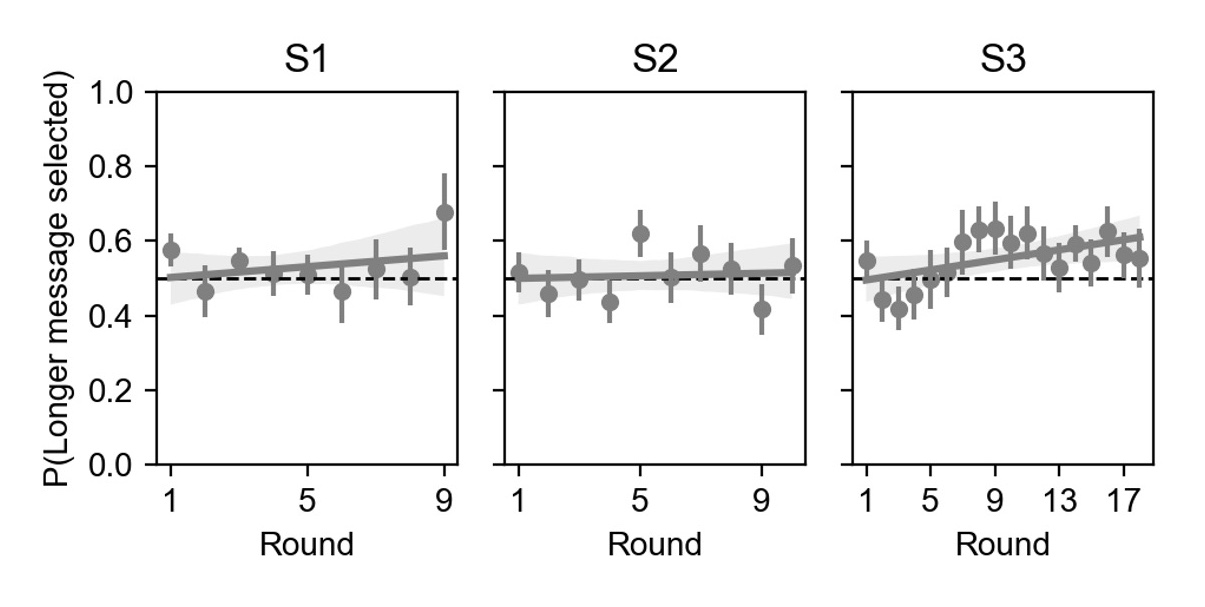}
    \caption{\textbf{Probability of candidates who wrote the longer message being selected in opaque conditions, conditional on selectors investing.} These probabilities did not go beyond 0.5, indicating selectors did not reveal selection biases toward candidates writing longer messages.}
    \label{fig:long_msg}
\end{figure}

\begin{figure}[H]
    \centering
    \includegraphics[width=1\linewidth]{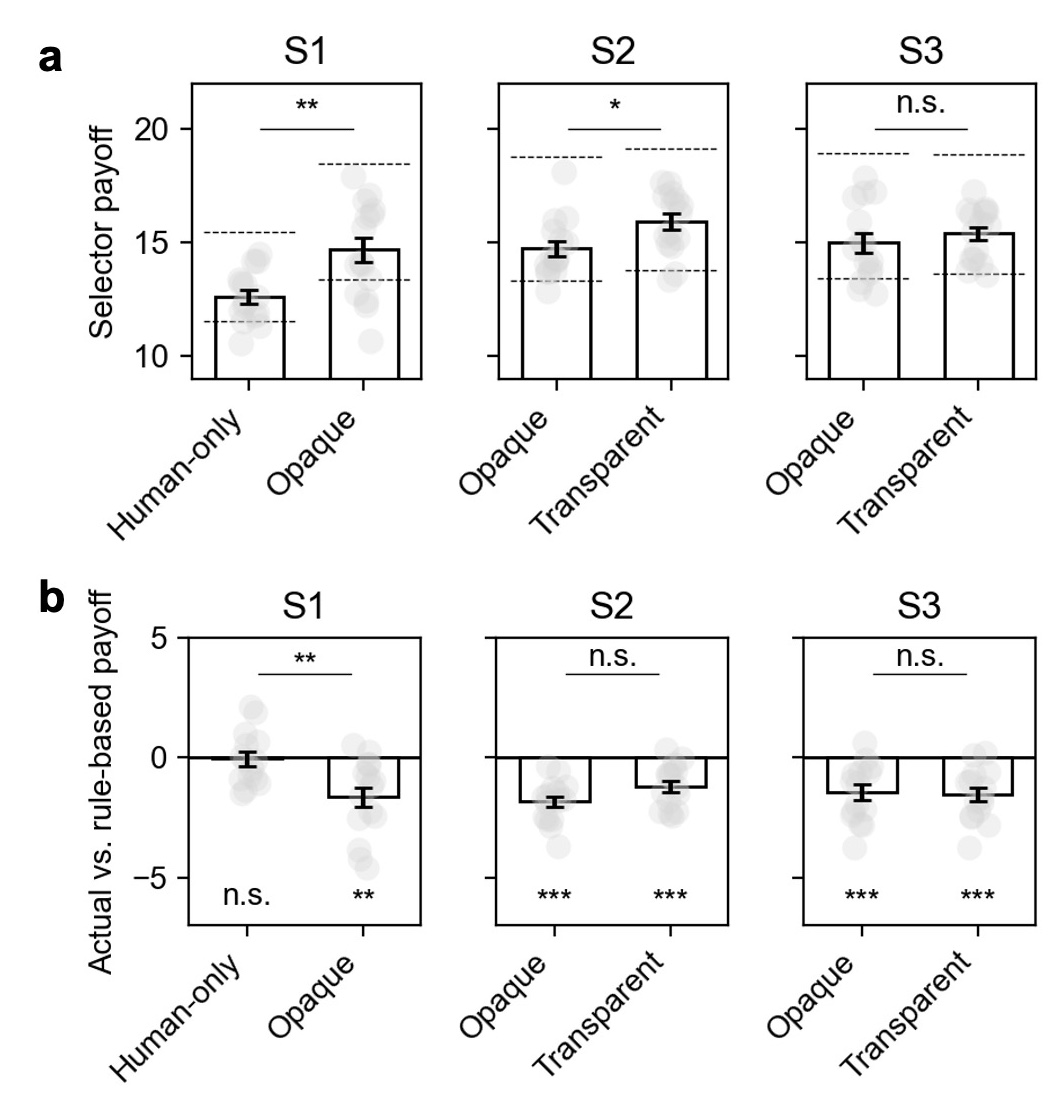}
    \caption{\textbf{Selector payoff.} \textbf{a}, Selectors' average payoff in each experimental condition. Lower and upper dashed lines indicate lower bounds (payoff of random choices) and upper bounds (maximum payoff selectors could have obtained given candidate returns) of selectors' payoff, respectively. In Study 1, selectors obtained higher payoffs in the opaque hybrid condition, because they were able to select bot candidates that on average returned much more points than human candidates. In Study 2, transparency helped selectors obtained higher payoff. However, this improvement was less pronounced in Study 3, probably because longer interaction also improved selectors' learning and decisions without transparency. \textbf{b}, Selectors' payoff versus the payoff that could be obtained by a rule-based selector who always selects the candidate writing the longer message. The rule-based selector outperformed human selectors in all hybrid conditions, but not in the human-only condition. The payoff gaps in hybrid conditions can partly be attributed to the miscalibration of selectors' belief in opaque conditions and the initial bias against bots in transparent conditions.}
    \label{fig:selector_payoff}
\end{figure}

\begin{figure}[H]
    \centering
    \includegraphics[width=1\linewidth]{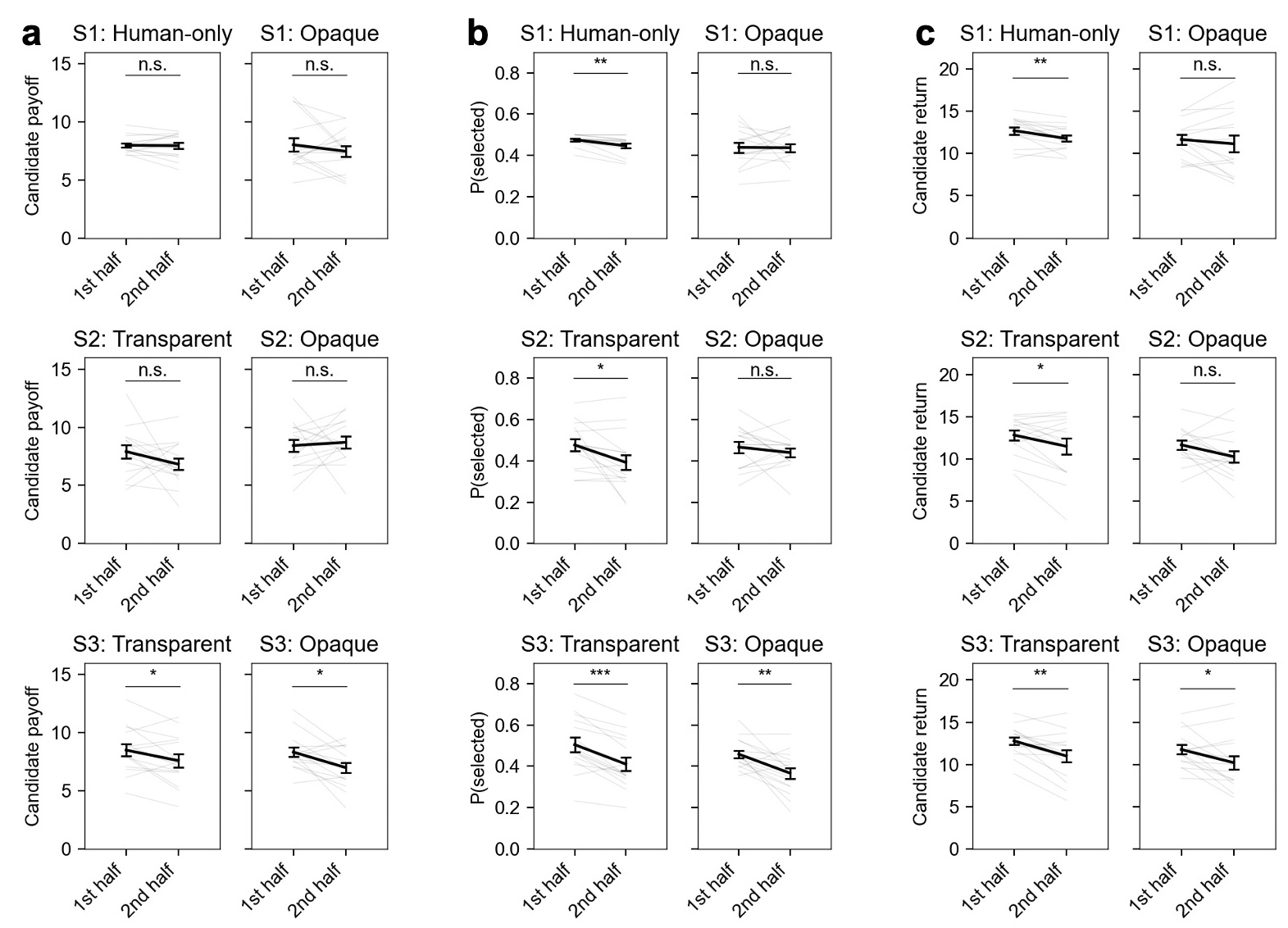}
    \caption{\textbf{Candidates' payoff (a), probability of being selected (b), and points returned (c) in the first and second halves of the game.} Candidates' payoffs were jointly determined by their likelihood of being selected and the amount they returned to selectors. In the human-only condition of Study 1 and the transparent condition of Study 2, candidates were selected less often in the second half, but their payoffs remained stable due to reduced returns. However, it is unclear whether lower returns led to fewer selections or vice versa. In Study 3, with longer interactions, the decline in selection outweighed the reduction in returned points, leading to decreased candidate payoffs in both transparent and opaque conditions.}
    \label{fig:candidate_payoff}
\end{figure}
\newpage


\begin{table}[ht]
\centering
\caption{Participant demographics across experimental conditions}
\label{tab:demo}
\begin{tabular}{@{}lccc@{}}
\toprule
\textbf{Condition} & \textbf{N} & \textbf{\# of females} & \textbf{Mean age ± s.d.} \\
\midrule
Study 1: human-only       & 225 & 135 & 34.90 ± 10.22 \\
Study 1: opaque           & 150 & 93  & 36.80 ± 12.51 \\
Study 2: opaque           & 150 & 79  & 36.02 ± 12.08\\
Study 2: transparent      & 150 & 86  & 35.53 ± 11.86 \\
Study 3: opaque           & 150 & 90  & 35.34 ± 10.94 \\
Study 3: transparent      & 150 & 84  & 34.44 ± 11.91 \\
\textbf{Total}              & \textbf{975} & \textbf{567} & \textbf{35.65 ± 13.00} \\
\bottomrule
\end{tabular}
\end{table}

\newpage

\section*{Supplementary Note 1: Text analyses on messages}

\begin{figure}[H]
    \centering
    \includegraphics[width=1\linewidth]{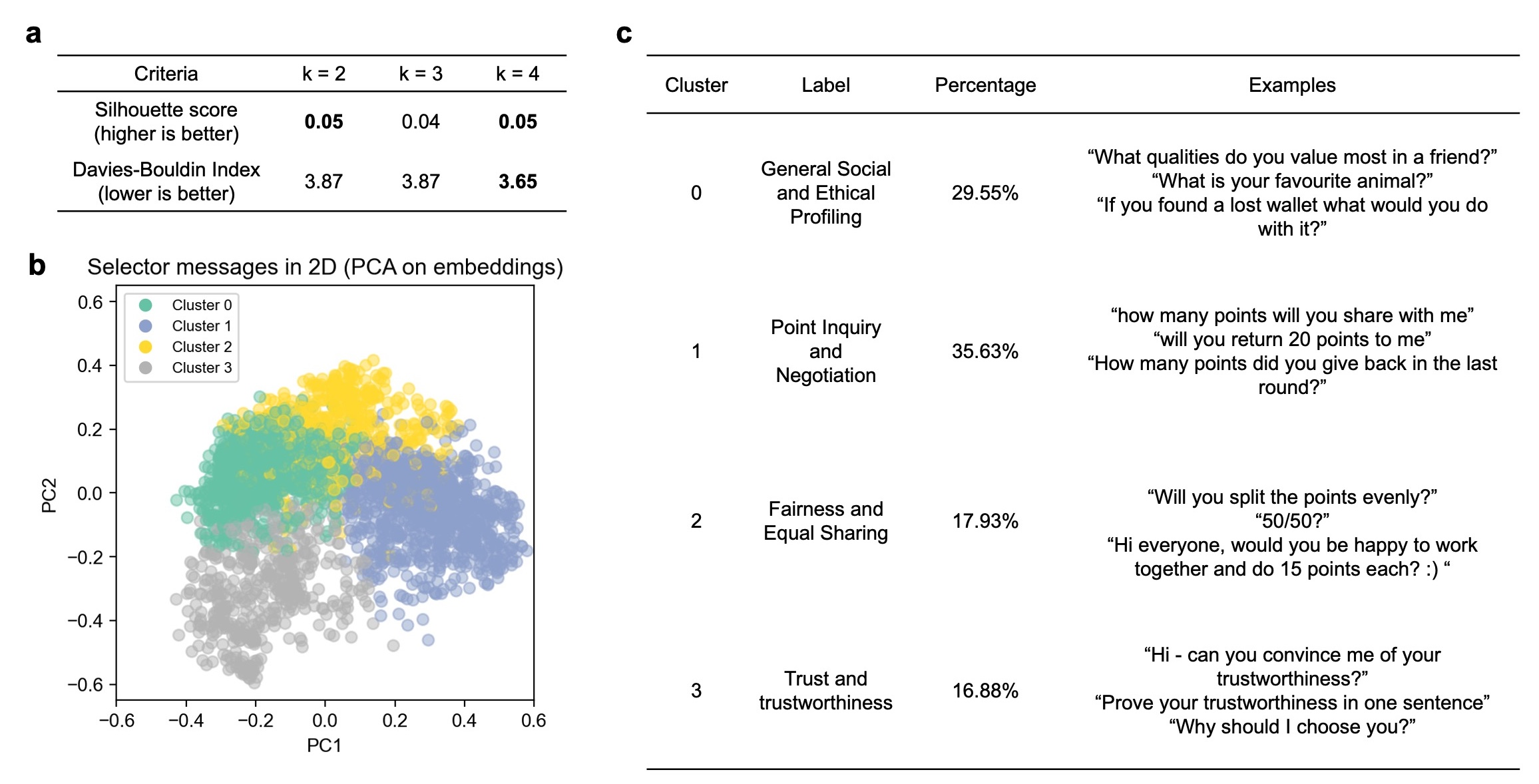}
    \caption{\textbf{Unsupervised clustering of selectors' messages.} \textbf{a}, To determine the number of clusters for $k$-means clustering on message embeddings, we evaluated models with $k = 2$, $3$, and $4$ using the Silhouette score and Davies–Bouldin index. Larger values of $k$ were excluded to maintain interpretability. Based on both metrics, we selected $k = 4$. \textbf{b}, Clusters were visualized by applying principal component analysis (PCA) to the message embeddings and projecting them onto the first two principal components. Each point represents a selector message. \textbf{c}, Manually assigned labels, proportions, and representative examples for each message cluster.}
    \label{fig:unsupervised}
\end{figure}

\begin{figure}[H]
    \centering
    \includegraphics[width=1\linewidth]{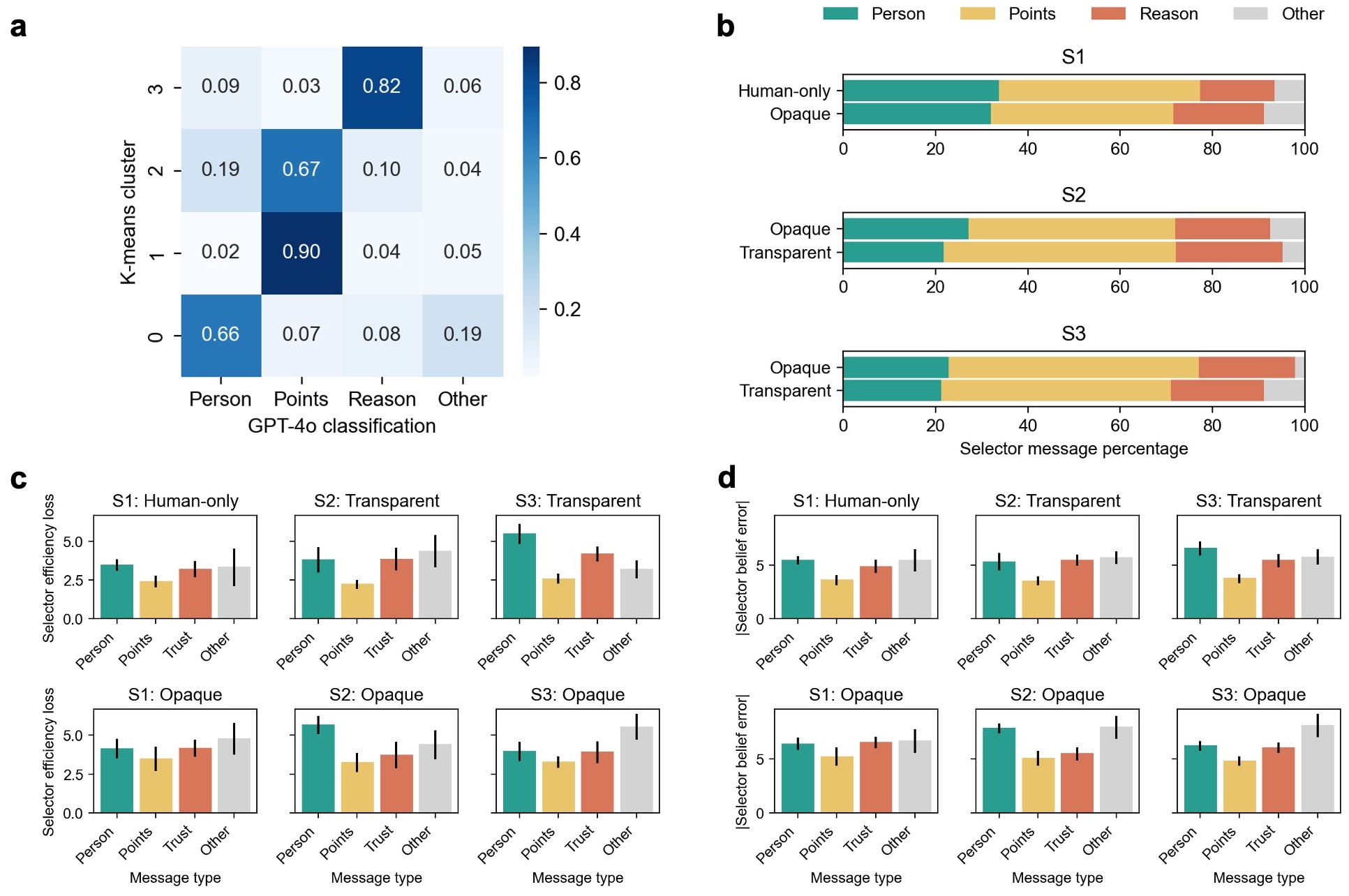}
    \caption{\textbf{Supervised classification of selectors' messages.} To validate the manual labels assigned to the clusters of selector messages derived from $k$-means clustering, we prompted GPT-4o to classify each message into one of four predefined categories: (1) questions about personal traits, (2) questions about points, (3) questions about reasons to select a candidate, or (4) other (see Supplementary Table~\ref{tab:selector_prompts} for prompt details). \textbf{a}, Heatmap showing the alignment between the unsupervised clustering and GPT-4o’s classification. Cluster 0 primarily comprised trait-related questions; Clusters 1 and 2 focused on points; Cluster 3 was mainly concerned with selection reasons. Values represent the conditional probability of GPT-4o assigning a message to a category, given its cluster derived from the $k$-means clustering. \textbf{b}, Proportions of message types (as classified by GPT-4o) across experimental conditions. \textbf{c}–\textbf{d}, To explore the relationship between message type and selector performance, we computed the average efficiency loss (\textbf{c}) and absolute belief error (\textbf{d}) for rounds involving different message types. Across all conditions, messages about points appear to be associated with best selector performance. Efficiency loss was defined as the gap between the selector’s actual payoff and the maximum possible payoff given candidates’ returns. Absolute belief error was the absolute difference between the selector’s belief and candidates' actual return.}
    \label{fig:supervised}
\end{figure}

\begin{figure}[H]
    \centering
    \includegraphics[width=1\linewidth]{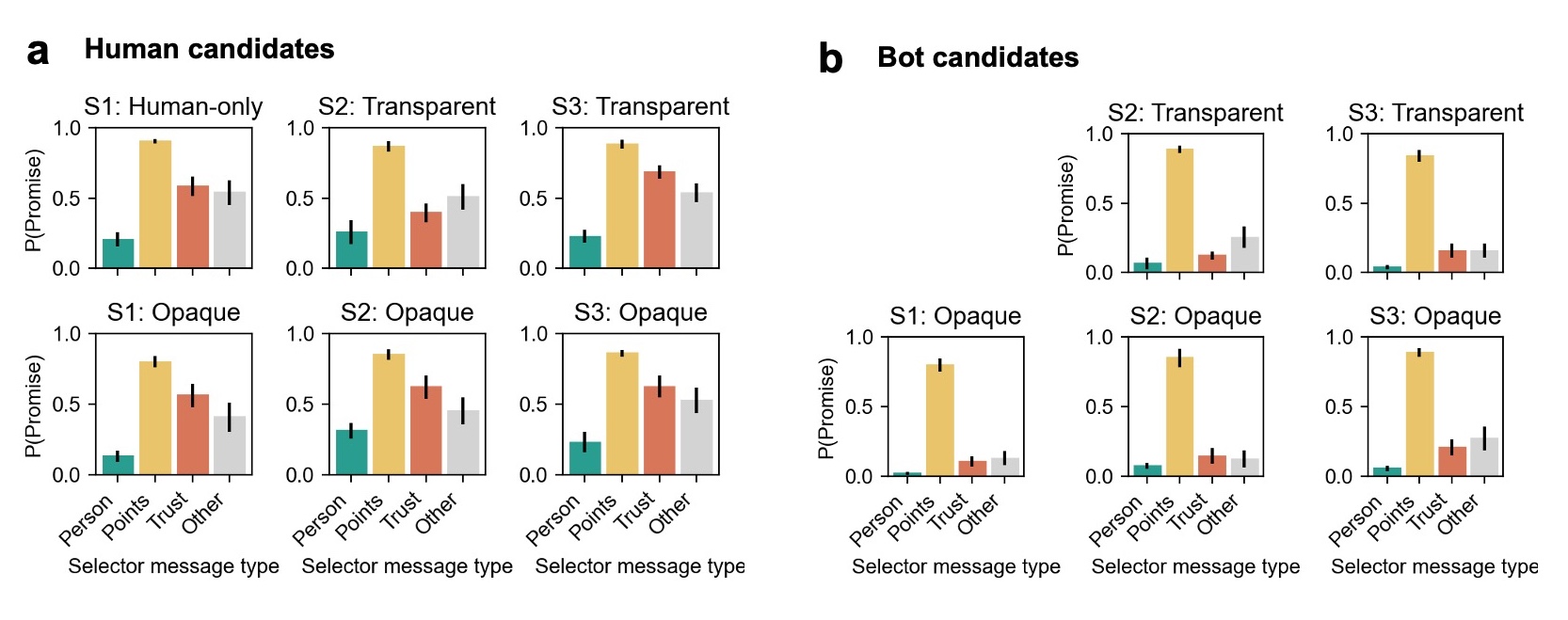}
    \caption{\textbf{Probability of candidates making promises conditional on selectors' message type (categorized by GPT-4o).} Across all conditions, both human (\textbf{a}) and bot (\textbf{b}) candidates made promises most frequently when selectors directly asked about points.}
    \label{fig:p_promise}
\end{figure}

\begin{figure}[H]
    \centering
    \includegraphics[width=1\linewidth]{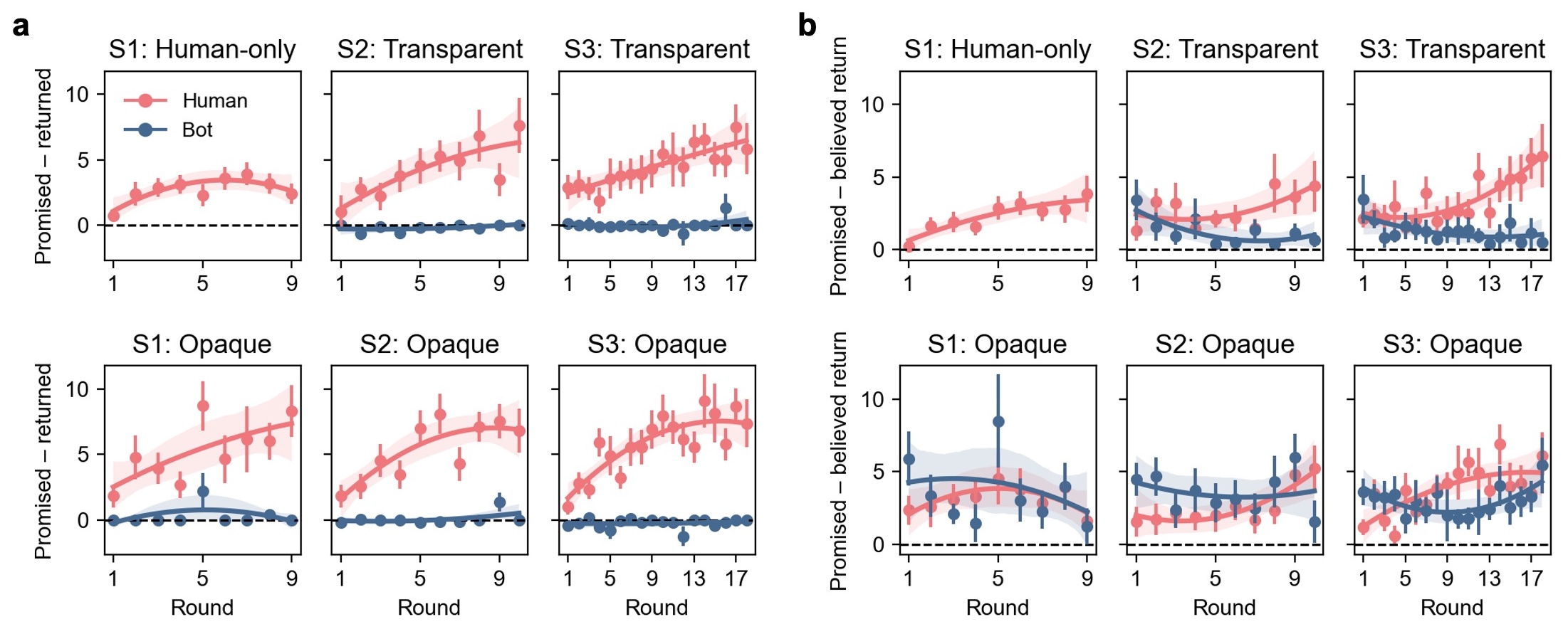}
    \caption{\textbf{Candidates' promised returns relative to their actual returns (a) and selectors' beliefs (b)}. 
    \textbf{a}, Bots consistently fulfilled their promises, whereas human candidates tended to return less than they promised, particularly in later rounds. Only cases where candidates made explicit promises were included in the analysis. 
    \textbf{b}, Selectors gradually learned that human candidates over-promised, as indicated by the positive difference between promised and believed returns. In contrast, only under transparent conditions did selectors recognize that bots were not like humans and reliably honored their promises, with believed returns converging toward the actual promises.}
    \label{fig:promise}
\end{figure}

To classify selectors' messages, we combined unsupervised and supervised methods to balance interpretability with reduced human bias (Feuerriegel et al., 2025). We first applied an unsupervised approach by embedding all selector messages pooled across all experiments using OpenAI’s \texttt{text-embedding-3-small} model and performing k-means clustering on the resulting embeddings to identify semantically similar groups (\ref{fig:unsupervised}). Clusters were then manually reviewed and assigned descriptive labels based on representative messages. To validate these labels, we used a supervised classification approach with GPT-4o. The model was prompted to categorize messages into four predefined types using category descriptions and illustrative examples (See Supplementary Tables \ref{tab:parameters} and \ref{tab:selector_prompts} for the parameters and prompts we used). This classification showed high agreement with the unsupervised clustering (\ref{fig:supervised}). For candidate messages, we employed GPT-4o to detect whether a promise was made and, if so, to extract the number of points promised for return (Supplementary Table \ref{tab:candidate_prompts}; \ref{fig:p_promise} and \ref{fig:promise}). 

\begin{table}[ht]
\caption{Parameters used for text analyses.}
\label{tab:parameters}%
\begin{tabular}{@{}lp{0.8\linewidth}@{}}
\toprule
Parameter & Value \\
\midrule
model & \texttt{"gpt-4o"} \\
max\_tokens & 200 \\
temperature & 0 \\
top\_p & 1 \\
response\_format & \texttt{\{"type": "json\_object"\}} \\
\botrule
\end{tabular}
\end{table}

\begin{table}[ht]
\caption{Prompts used for classifying selectors' messages.}
\label{tab:selector_prompts}%
\begin{tabular}{@{}lp{0.8\linewidth}@{}}
\toprule
Variable & Value \\
\midrule
system message & Your task is to categorize messages from an experiment on partner selection games. The game rules are as follows: There is one selector and two candidates in one game. The selector is endowed with 10 points and can decide whether to select one of the candidates to be an investment partner or to keep all the points for themself. If the selector selects one of the candidates as the partner, the 10 points will all be transferred to the selected candidate and tripled. The selected candidate can decide how many points out of 30 they want to return to the selector. The unselected candidate will receive no points. Before making their decision, the selector can send a question to both candidates to probe their trustworthiness, and the candidates can reply to convince the selector. \\ 
 & The user will provide the selector's question. Your task is to categorize the message into one of the following 4 categories: \\
 & 1. general questions about candidates' personal traits, including values, preferences, personalities, and capabilities. These may be related to personal opinions, experiences, or habits. Examples: What is your favorite color; What do you think of green energy; Do you take risks? \\
 & 2. questions or proposals about the points candidates will return. Examples: How many points will you return; Will you split the points equally; 50/50? \\
 & 3. asking candidates to prove their trustworthiness or provide reasons to select them. Examples: Convince me that you are trustworthy; Why should I choose you? \\
 & 4. other categories or cannot be categorized into the above categories.\\
 & You respond in the JSON format with the key: 'category'. The values should be integers between 1 and 4, corresponding to the question category. \\
user message & Selector's question: \texttt{<SELECTOR MESSAGE>}\footnotemark[1] \\
\botrule
\end{tabular}
\footnotetext[1]{\texttt{<SELECTOR MESSAGE>} was replaced by the actual message from the selector. Similarly below.}
\end{table}

\begin{table}[ht]
\caption{Prompts used for encoding candidates' messages.}
\label{tab:candidate_prompts}%
\begin{tabular}{@{}lp{0.8\linewidth}@{}}
\toprule
Variable & Value \\
\midrule
system message & Two players are playing partner selection games. There is one selector and two candidates in one game. The selector is endowed with 10 points and can decide whether to select one of the candidates to be an investment partner or to keep all the points for themself. If the selector selects one of the candidates as the partner, the 10 points will all be transferred to the selected candidate and tripled. The selected candidate can decide how many points out of 30 they want to return to the selector. The unselected candidate will get no points. Before the selector decides, they can send a question to both candidates to probe their trustworthiness, and the candidates can reply to convince the selector. \\
 & The user will provide the selector's question together with the replies from the two candidates. You determine for each of the candidate, how many points did they promise to return to the selector. You respond in the JSON format with two keys: 'promise\_a' and 'promise\_b', corresponding to the promise made by candidate A and candidate B, respectively. The values should be integers between -1 and 30, where -1 means the candidate did not make a specific promise. \\
user message & Selector: \texttt{<SELECTOR MESSAGE>}\\
 & Candidate A: \texttt{<CANDIDATE A MESSAGE>}\\
 & Candidate B: \texttt{<CANDIDATE B MESSAGE>}\\
\botrule
\end{tabular}
\end{table}

\clearpage

\section*{Supplementary Note 2: Computational modeling of selectors' belief updating}

To better understand how selectors formed and updated beliefs about the returns of human and bot candidates and how this process was influenced by identity transparency, we developed computational models that characterize round-by-round belief updates based on the feedback of candidates' returns.

Our models are based on the Rescorla–Wagner (RW) reinforcement learning algorithm, widely used to model human social learning \cite{sutton_reinforcement_1998, olsson_neural_2020, jiang_neurocomputational_2023}. In this framework, selectors update their beliefs about candidate returns based on \textit{prediction errors}: the difference between actual and expected returns, weighted by a \textit{learning rate}.

\begin{figure}[H]
    \centering
    \includegraphics[width=1\linewidth]{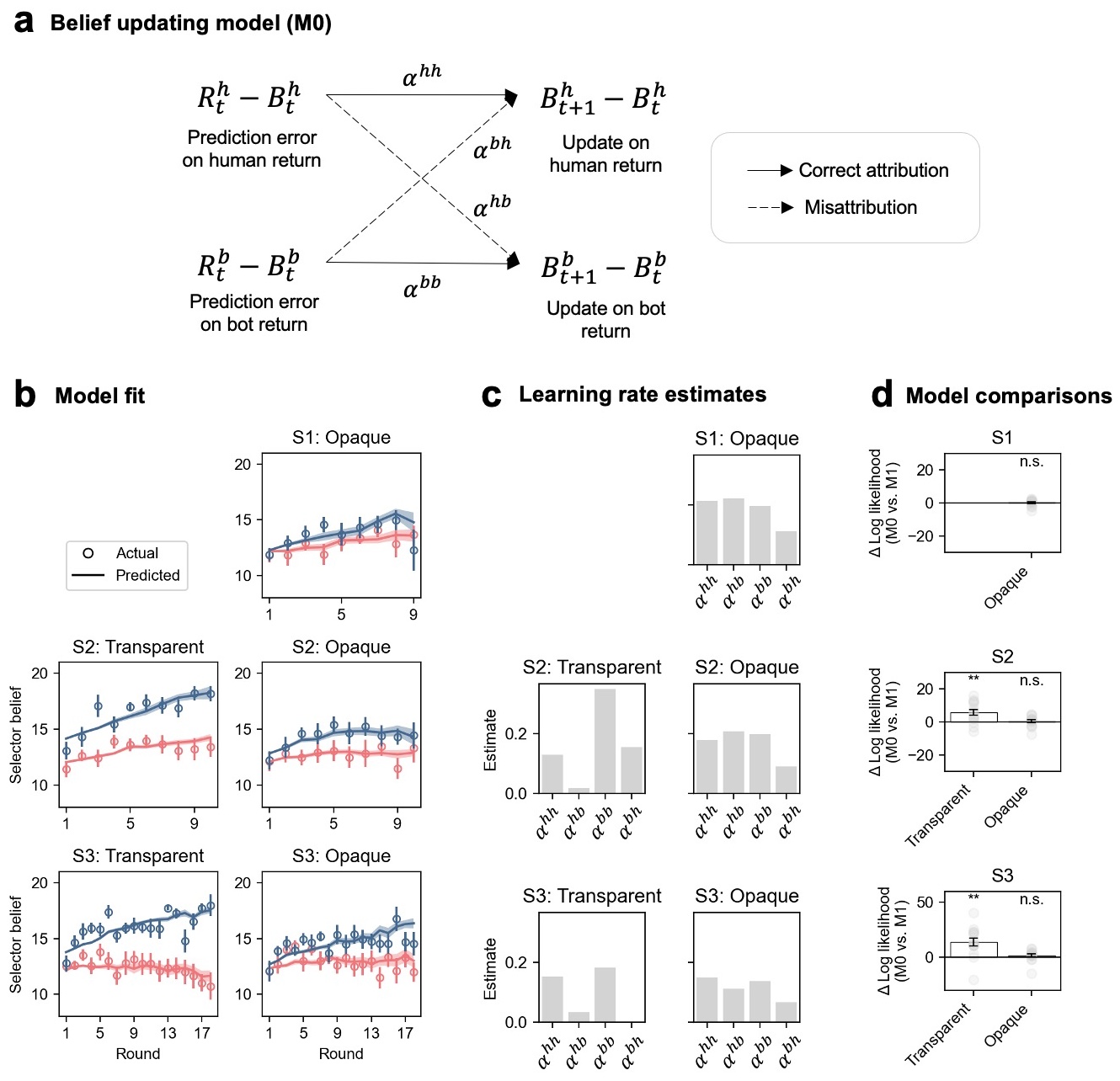}
    \caption{\textbf{Computational modeling of selectors’ belief updating.} \textbf{a}, Illustration of correct and incorrect attribution paths in our full belief-updating model (M0). \textbf{b}, Predictions from the best-fitting full model closely matched selectors’ reported beliefs. Dots and error bars indicate mean and intergroup s.e.m. of reported beliefs (replicated from \ref{fig:belief}); lines and shaded areas show the mean and s.e.m. of out-of-sample predictions from the model. \textbf{c}, Learning rates estimated from the full model, fit to pooled data by condition. Nonzero $\alpha_{hb}$ and $\alpha_{bh}$ indicate misattribution across candidate types. In transparent conditions, $\alpha_{hb}$ was near zero, suggesting that prediction errors from human candidates were hardly misattributed to bots. However, in Study 2 (but not S3), bot-triggered prediction errors were partially misattributed to humans ($\alpha_{bh} > 0$), consistent with selectors' mild but nonsignificant overestimation of human returns (\ref{fig:belief}). In opaque conditions, prediction errors from human candidates influenced beliefs about both humans and bots, contributing to selectors' underestimation of bot returns. Prediction errors from bots were also partially misattributed to humans, contributing to the overestimation of human returns. \textbf{d}, Model comparison based on out-of-sample log-likelihood. The full model outperformed the reduced model only in the transparent conditions but not opaque conditions, suggesting that selectors only formed separate beliefs about two types of candidates when their identities were disclosed. Error bars represent intergroup s.e.m. $^{**}p<0.01$.}
    \label{fig:model}
\end{figure}

The \textbf{full model (M0)} extends the classical RW model by allowing both correct and incorrect attribution of feedback (see \ref{fig:model} for a visual illustration). That is, prediction errors from selecting a human (or bot) candidate can influence beliefs about both human and bot candidates, with separate learning rates. For simplicity, we assumed selectors only updated their beliefs when they selected a candidate; no update occurred when they chose to keep the points.

If a \textit{human candidate} was selected in round $t$, beliefs were updated as:
\begin{equation}
B_{t+1}^h = B_t^h + \alpha^{hh}(R_t^h - B_t^h)
\end{equation}
\begin{equation}
B_{t+1}^b = B_t^b + \alpha^{hb}(R_t^h - B_t^h)
\end{equation}

If a \textit{bot candidate} was selected:
\begin{equation}
B_{t+1}^h = B_t^h + \alpha^{bh}(R_t^b - B_t^b)
\end{equation}
\begin{equation}
B_{t+1}^b = B_t^b + \alpha^{bb}(R_t^b - B_t^b)
\end{equation}

Here, $B_t^h$ and $B_t^b$ denote the selector’s belief at round $t$ about human and bot candidates' returns, respectively; $R_t^h$ and $R_t^b$ are the actual returns observed. The $\alpha$ parameters represent learning rates, allowing for asymmetric and cross-type learning. Non-zero $\alpha_{hb}$ and $\alpha_{hb}$ estimates indicate across-type misattribution.

To account for noise in reported beliefs, we assumed for all models that each belief report at round $t$ was sampled from a normal distribution:
\[
\hat{B}_t \sim \mathcal{N}(B_t, \sigma^2)
\]
where $B_t = B_t^h$ or $B_t^b$ depending on the candidate type.

To verify whether identity transparency helped selectors form separate beliefs about human and bot candidates, we also compared the full model with a \textbf{reduced model (M1)} that assumes selectors could not differentiate two candidate types and formed a single belief about candidate returns ($B_t^h = B_t^b$) by setting a shared learning rate ($\alpha^{hh} = \alpha^{hb} = \alpha^{bh} = \alpha^{bb}$). If the selectors formed separate beliefs about human and bot candidates, the full model should outperform the reduced model in predicting selectors' belief dynamics.

The free parameters—learning rates $\alpha$, initial beliefs $B_0^h$ and $B_0^b$, and reporting noise $\sigma$—were estimated by maximizing the likelihood of the observed belief reports. Model comparison was performed using leave-one-group-out cross-validation.

\end{appendices}


\end{document}